\documentclass{emulateapj}
\usepackage{graphicx,graphics,amsmath}

\def\Rs{R_{\odot}}

\def\mc{meridional circulation}
\def\ftd{flux transport dynamo}
\def\pa{\partial}
\def\er{\mbox{erf}}
\newcommand{\ov}{\overline}
\begin{document}

\title{Is a deep one-cell meridional circulation essential for the flux
transport solar dynamo?}
\author{Gopal Hazra$^{1,2}$, Bidya Binay Karak$^{1,3}$, and Arnab Rai Choudhuri$^1$}
\affil{$^1$Department of Physics, Indian Institute of Science, Bangalore 560012, India\\
$^2$Indian Institute of Astrophysics, Bangalore 560034, India\\
$^3$Nordita KTH Royal Institute of Technology and Stockholm University, Roslagstullsbacken 23, SE-106 91 Stockholm, Sweden}

\begin{abstract}

The solar activity cycle is successfully modeled by the flux transport
dynamo, in which the \mc\ of the Sun plays an important role.  Most of the
kinematic dynamo simulations assume a one-cell structure of the \mc\ within the
convection zone, with the equatorward return flow at its bottom.
In view of the recent claims that the return flow occurs
at a much shallower depth, we explore whether a \mc\ with such a shallow
return flow can still retain the attractive features of the flux transport
dynamo (such as a proper butterfly diagram, the proper phase relation between
the toroidal and poloidal fields). We consider additional cells of the \mc\
below the shallow return flow---both the case of multiple cells radially
stacked above one another and the case of more complicated cell patterns.
As long as there is an equatorward flow in low latitudes at the bottom of
the convection zone, we find that the solar behavior is approximately
reproduced.  However, if there is either no flow or a poleward flow at the
bottom of the convection zone, then we cannot reproduce solar behavior.
%although this
%conclusion may change if there is suitable turbulent pumping in the equatorward direction.
On making the turbulent diffusivity low, we still find periodic behavior,
although the period of the cycle becomes unrealistically large.  Also, with
a low diffusivity, we do not get the observed correlation between the polar
field at the sunspot minimum and the strength of the next cycle, which is
reproduced when diffusivity is high.  On introducing radially downward
pumping, we get a more reasonable period and more solar-like behavior even
with low diffusivity.

\end{abstract}
\email{ghazra@physics.iisc.ernet.in}
\maketitle
\section{Introduction}

The most extensively studied theoretical model of the solar activity cycle
in the last few years is the flux transport dynamo model, originally proposed
in the 1990s (Wang et al.\ 1991; Choudhuri et al.\ 1995;
Durney 1995) and recently reviewed by several authors (Charbonneau 2010; Choudhuri 2011).
This model had a remarkable success in explaining various aspects of the solar
cycle and its irregularities.  However, in spite of its success, doubts are often
expressed if this success is merely accidental or if this is the really correct
model. Basically this hinges on the question whether the various assumptions used
in this model are correct.

Let us consider the crucial assumptions of the model.  The toroidal magnetic
field is assumed to be produced from the poloidal field
by the differential rotation that is mapped by
helioseismology, leaving no scope for any doubts.  This toroidal field rises
due to magnetic buoyancy to the solar surface, where the poloidal magnetic field
is produced from it by the Babcock--Leighton mechanism
(Babcock 1961; Leighton 1964), for which there is now strong observational support
(Dasi-Espuig et al.\ 2010; Kitchatinov \& Olemskoy 2011a; Mu\~noz-Jaramillo et al.\ 2013). It is true that magnetic
buoyancy and the Babcock--Leighton mechanism are inherently three-dimensional 
processes and their representation in a two-dimensional kinematic model can
never be fully realistic (Mu\~noz-Jaramillo et al.\ 2010; Yeates \& Mu\~noz-Jaramillo 2013). There are also considerable
uncertainties in the values of some parameters, such as the turbulent diffusivity
inside the convection zone. The Boulder group (Dikpati et al.\ 2004) and the
Bangalore group (Chatterjee et al.\ 2004) used values of turbulent
diffusivity which differ by a factor of about 50. The higher diffusivity used
by the Bangalore group has now got strong support due to the success in explaining
various aspects of the observational data (Chatterjee et al.\ 2004; Chatterjee
\& Choudhuri 2006; Jiang et al.\ 2007; Goel \& Choudhuri 2009;
Hotta \& Yokoyama 2010; Karak 2010; Karak \& Choudhuri 2011, 2012, 2013; Choudhuri \&
Karak 2009, 2012; Miesch et al.\ 2012; Mu\~noz-Jaramillo et al.\ 2013).  Because 
of these uncertainties in the treatment of the Babcock--Leighton
mechanism and in the value of turbulent diffusivity, it is necessary to interpret
the results of the flux transport dynamo model with some degree of caution.
However, these uncertainties do not invalidate the model.  After all, different
treatments of the Babcock--Leighton mechanism and a range of values for the
turbulent diffusivity seem to give qualitatively similar results.  The only
other important ingredient of the flux transport dynamo model is the meridional
circulation.  Because the nature of the \mc\ in the deeper layers of the convection
zone is not yet observationally established, the main source of doubt about the
flux transport dynamo model at the present time concerns the question of whether
the Sun really has the kind of \mc\ which is assumed in the flux transport dynamo
models.

Let us consider the role of the \mc\ in \ftd\ models. In order for sunspots to
form at lower and lower latitudes with the progress of the cycle, a condition known as
the Parker--Yoshimura sign rule was expected to be satisfied (Parker 1955; Yoshimura 1975;
see also Choudhuri 1998, Section 16.6).  According to this condition, $\alpha \, \pa \Omega/ \pa r$ 
has to be negative in the northern hemisphere.  It
follows from observations of solar surface that $\alpha$ corresponding to the
Babcock--Leighton mechanism is positive in the northern hemisphere.  Helioseismology
shows that $\pa \Omega/ \pa r$ is also positive in the lower latitudes where
sunspots are seen (except in a shear layer just below the solar surface). 
So clearly the Parker--Yoshimura sign rule is not satisfied
and it may be expected that the dynamo wave will propagate in the poleward
direction, contrary to observations.  Choudhuri et al.\ (1995)
showed that an equatorward meridional circulation at the bottom of the convection
zone can overcome the Parker--Yoshimura sign rule and make the dynamo wave
propagate in the correct direction.  This is the main role of the \mc\ in the
\ftd\ models.  The second role of the \mc\ is that the poleward \mc\ near the
solar surface advects the poloidal field poleward, as seen in the observations.
In dynamo models with low turbulent diffusivity, the \mc\ has a third important
role.  It brings the poloidal field created near the surface to the bottom of
the convection zone where the strong differential rotation can act on it to
create the toroidal field.  In dynamo models with high turbulent diffusivity,
however, the poloidal field can diffuse from the surface to the bottom of the
convection zone in typically about 5 yr and this third role of the \mc\ is 
redundant (Jiang et al.\ 2007). If there is radial pumping
as suggested by some authors (Karak \& Nandy 2012), then that further
eliminates the role of \mc\ for bringing the poloidal field to the bottom
of the convection zone. 
Because we will be using a higher value of turbulent diffusivity in many of our calculations, the twin roles of the \mc\ in our model 
will be the equatorward advection
of the toroidal field at the bottom of the convection zone and the poleward
advection of the poloidal field near the surface.

The simplest kind of \mc\ assumed in most of the 
theoretical models consists of one cell
encompassing a hemisphere of the convection zone, with a poleward flow in the
upper layers and an equatorward flow in the lower layers. This kind of \mc\
successfully plays the twin roles expected of it in a \ftd\ model. 
Observations show a poleward \mc\ near the surface, so there is absolutely no
doubt that this part of the \mc\ advects the poloidal field poleward. The
only remaining question is whether the cell of the \mc\ really penetrates to the 
bottom of the convection zone where the equatorward flow branch has to be
located for the equatorward advection of the toroidal field. Early helioseismic
investigations going to a depth of $0.85 R_{\odot}$ could not find any evidence
of the equatorward return flow until that depth (Giles et al.\ 1997; Braun \& 
Fan 1998). However, recently Hathaway (2012), assuming that the supergranules
are advected by the \mc, analyzed the observational data to conclude that the
return flow occurs at depths as shallow as 50--70 Mm. 
Zhao et al.\ (2013) also claim on the basis of their helioseismic inversion
that the equatorward return flow
exists between radii $0.82 R_{\odot}$ and $0.91 R_{\odot}$. On the other 
hand, Schad et al.\ (2013) from the study of global helioseismic analysis 
find the indication of multi-cell meridional circulation in the whole 
convection zone. If these results
are supported by other independent groups and 
really turn out to be true, then the very important question is whether the
attractive aspects of the present \ftd\ models can be retained with such a
\mc. In this paper, we explore whether additional cells of \mc\ below the
shallow return flow can help us solve the problem.

So far only a few theoretical studies of the \ftd\ with a \mc\ more 
complicated than a single cell have been carried out.  The effects of two
cells in the latitudinal direction have been considered by Dikpati et al.\
(2004) and Bonanno et al.\ (2005). However, we now want to consider a more
complicated structure of the \mc\ in the radial direction, including the
possibility of multiple
cells in the radial direction. Such a study was first carried out by Jouve
\& Brun (2007).  In their calculations, they always had poleward meridional
circulation at the bottom of the convection zone in the lower latitudes
where sunspots are seen.  They were able to get periodic solutions, but
the butterfly diagrams were always in the wrong sense, implying poleward
migration of the toroidal field.  They concluded that ``the resulting butterfly 
diagram and phase relationship between the toroidal and
poloidal fields are affected to a point where it is unlikely that such 
multicellular meridional flows persist for a long period of time
inside the Sun, without having to reconsider the model itself''(Jouve \& Brun 2007, p. 239). If this
conclusion was generally true for any \mc\ more complicated than the
simple single-cell circulation, then the results of Hathaway (2012),
Schad et al. (2013) and Zhao et al.\ (2013), if supported by independent investigations by
other groups, would indeed be bad news for
\ftd\ models.  Guerrero \& de Gouveia dal Pino (2008) considered a single
cell confined to the upper layers of the convection zone.  On introducing
strong radial and latitudinal pumping, they were able to get correct
butterfly diagrams.  However, whether such equatorward latitudinal pumping
really exists to give the right kind of butterfly diagram is highly questionable.
Another recent attempt of saving the \ftd\ has been made by Pipin \&
Kosovichev (2013), who use the near-surface shear layer found in helioseismology
and an equatorward return flow of \mc\ just below it. Since $\pa \Omega/ \pa r$ is 
negative within this shear layer, such a dynamo would have equatorward
propagation according to the Parker--Yoshimura sign rule
even in the absence of an equatorward \mc\ in the right place.
However, we are unable to accept this model of Pipin \& Kosovichev (2013) as a 
satisfactory model of the solar cycle for the following
reasons. It is known for a long time that magnetic buoyancy is particularly
destabilizing in the upper layers of the convection zone and it is impossible
to store magnetic fields generated there for sufficient time for
dynamo amplification (Parker 1975; Moreno-Insertis 1983). Also, the
scenario that the toroidal field is generated within the tachocline and then
parts of it rise to produce active regions can explain many aspects of active
regions including Joy's law rather elegantly (Choudhuri 1989; D'Silva \& Choudhuri 1993;
Fan et al.\ 1993; Caligari et al.\ 1995). We find no compelling
reason to discard the scenario that the toroidal magnetic field is produced
in the tachocline where magnetic buoyancy is suppressed in the regions of
sub-adiabatic temperature gradient (Moreno-Insertis et al.\ 1992). 

The main aim of the present paper is to address the question whether a \mc\
with a return flow at a relatively shallow depth would allow us to retain the
attractive features of the \ftd, without introducing such uncertain
assumptions as strong equatorward pumping and without abandoning the scenario in which the toroidal
field is generated and stored in the tachocline from where it rises to produce
active regions. If there is a return flow at a shallow depth and there are
no flows underneath it, then we find that the solar cycle cannot be
modeled properly with such a flow.  However, if there are additional
cells of \mc\ below the shallow return flow,
we find that we can retain most of the attractive features
of the \ftd\ model as long as there is a layer of equatorward flow in low latitudes
at the bottom of the convection zone.  The existence of such an equatorward
flow at the bottom of the convection zone is consistent with the findings 
of Zhao et al.\ (2013), who are unable to extend their inversion below
$0.75 R_{\odot}$ using their limited data set. Since our knowledge of the
\mc\ either from observational or theoretical considerations is very limited,
in this paper we take the \mc\ as a free parameter that can be assumed
to have any form involving multiple cells and explore the dynamo problem
with different kinds of \mc.
 
We discuss the mathematical formulation of our dynamo model in Section~2. Then in
Section~3 we present our results for several cells of \mc\ in the radial direction,
whereas Section~4 presents results for more complicated \mc\ with multiple cells
in both radial and latitudinal directions. Whether the results get modified
for low turbulent diffusivity will be discussed in Section~5.  The effect of turbulent
pumping will be discussed in Section~6.  Finally we summarize our conclusions in
Section~7.
%^^^^^^^^^^^^^^^ Mathematical formulation ^^^^^^^^^^^^^^^^^^^^^^^^^^^^^^^^^^^
\section{Mathematical formulation}

In the two-dimensional kinematic flux transport dynamo model,
we represent the magnetic field as
\begin{equation}
{\bf B} = B \hat{\bf e}_{\phi} + \nabla \times (A \hat{\bf e}_{\phi}),
\end{equation}
where $B (r, \theta)$ and $A(r, \theta)$ respectively correspond to the 
toroidal and poloidal components which satisfy the following equations:
\begin{equation}
\label{eq:Aeq}
\frac{\partial A}{\partial t} + \frac{1}{s}({\bf v}.\nabla)(s A)
= \eta_{p} \left( \nabla^2 - \frac{1}{s^2} \right) A + S(r, \theta, t),
\end{equation}

\begin{eqnarray}
\label{eq:Beq}
\frac{\partial B}{\partial t}
+ \frac{1}{r} \left[ \frac{\partial}{\partial r}
(r v_r B) + \frac{\partial}{\partial \theta}(v_{\theta} B) \right]
= \eta_{t} \left( \nabla^2 - \frac{1}{s^2} \right) B \nonumber \\
+ s({\bf B}_p.{\bf \nabla})\Omega + \frac{1}{r}\frac{d\eta_t}{dr}\frac{\partial{(rB)}}{\partial{r}},~~~~~~~~~~~~~~~~~~~~
\end{eqnarray}\\
where $s = r \sin \theta$.  Here ${\bf v}$ is velocity of the meridional flow, $\Omega$
is the internal angular velocity of the Sun, $\eta_p$ and $\eta_t$ are
turbulent diffusivities
and $S(r, \theta, t)$ is the coefficient which describes the generation
of poloidal field at the solar surface from the decay of bipolar
sunspots. These equations have to be solved with the boundary conditions
$A=B=0$ at $\theta = 0, \pi$, whereas at the top boundary $B=0$ and $A$
matches a potential field above (Dikpati \& Choudhuri 1995).
The bottom boundary condition does not affect the solutions as long
as the bottom of the integration region is taken sufficiently below
the bottom of the convection zone. 
Once the parameters $\Omega$, $\eta_p$, $\eta_t$, ${\bf v}$ and
$S(r, \theta, t)$ are specified, Equation (2) and (3) can be solved
with the code {\em Surya} to obtain the behavior of the dynamo
(Nandy \& Choudhuri 2002; Chatterjee et al.\ 2004). Chatterjee et al.\ (2004)
present a detailed discussion how the parameters were specified in their
simulations. However, Karak (2010) made some small changes in the parameters.
In the calculations in this paper, we use the $\Omega$, $\eta_p$ and $\eta_t$ as
Karak (2010), except that Section~5 and Section~6 present some discussion with different diffusivities
which will be explained in Section~5.  In this paper we
carry out dynamo simulations with different kinds of meridional circulation ${\bf v}$.
Before coming to the meridional circulation, let us describe how we specify the
poloidal source term $S(r, \theta, t)$.

The effects of the magnetic buoyancy and the Babcock--Leighton mechanism have
to be incorporated by suitably specifying the poloidal source term $S(r, \theta, t)$.
There are two widely used procedures of specifying magnetic buoyancy.  In the
first procedure, whenever the toroidal field $B$ at the bottom of the convection
zone crosses a critical value, a part of it is brought to the solar surface.
In the second procedure, the Babcock--Leighton coefficient $\alpha$ in the 
source term multiplies the toroidal magnetic field at the bottom of the convection
zone rather than the local toroidal field.  Although the two procedures with all the
other parameters kept the same do not give identical results (Choudhuri et al.\ 2005),
both the procedures reproduce the qualitative behaviors of the solar cycle.  Since
we believe the first procedure to be more realistic, we had used it in the majority
of calculations from our group (Chatterjee et al.\ 2004; Choudhuri et al.\ 2007;
Karak 2010). However, it sometimes becomes difficult to introduce this procedure
in a stable way when the meridional circulation is made very complicated. Because
we are studying the behavior of the dynamo with various
complicated meridional circulations, we have opted for the second procedure.  We
specify the poloidal source term in Equation (2) in the following way:
\begin{equation}
 S(r, \theta; B) = \frac{\alpha(r,\theta)}{1+(\ov B(r_t,\theta)/B_0)^2} \ov B(r_t,\theta),
\label{source}
\end{equation}
where $\ov B(r_t,\theta)$ is the value of the toroidal field at latitude $\theta$
averaged over the tachocline from $r = 0.685\Rs$ to $r=0.715\Rs$. We take
\begin{eqnarray}
\alpha(r,\theta)=\frac{\alpha_0}{4}\left[1+\mathrm{erf}\left(\frac{r-0.95\Rs}{0.05\Rs}\right)\right]\left[1-\mathrm{erf}\left(\frac{r-\Rs}{0.01\Rs}\right)\right]\nonumber\\
\times \sin\theta\cos\theta\left[\frac{1}{1+e^{-30(\theta-\pi/4)}}\right]~~~~~~~~~~~~
\label{alpha}
\end{eqnarray}
Note that the last factor in Equation (5) suppresses $\alpha$ in the higher latitudes and constrains
the butterfly diagram from extending to very high latitudes.  We are following
many previous authors who also suppressed $\alpha$ in high
latitudes by such means (Mu\~noz-Jaramillo et al.\ 2010; Hotta \& Yokoyama 2010). Since this suppression
of $\alpha$ is not based on a clear physical reason, we have not used such a suppression of $\alpha$
in the previous calculations from our group using the first procedure of treating magnetic buoyancy
outlined above.  However, on treating magnetic buoyancy by the second procedure, flux eruptions tend
to occur at higher latitudes (Choudhuri et al.\ 2005) and it becomes necessary to suppress eruptions
at high latitudes to get more reasonable butterfly diagrams.
For calculations presented in Section~3 and 4 using high diffusivity, we use $\alpha_0=8.0$ m s$^{-1}$. 
In the low diffusivity case presented in Section~5, we use a lower value $\alpha_0=0.5$ m s$^{-1}$. When 
we include the effect of radial turbulent pumping in Section~6 we use $\alpha_0=0.1$ m s$^{-1}$.
Note that the parameter $B_0$ in Equation (4) introduces the only nonlinearity in the problem and determines the amplitude of
the magnetic field.  We will later present magnetic fields in units of $B_0$.

Below we discuss how the meridional circulation is prescribed.
The meridional circulation is always defined in terms of stream function $\psi$ which is given by
\begin{equation}
\rho {\bf v} = \nabla \times [\psi (r, \theta) {\bf e}_{\phi}],
\end{equation}
with the density profile given by
\begin{equation}
\rho = C \left( \frac{R_\odot}{r} - 0.95 \right)^{3/2}, \label{rho}
\end{equation}
We can generate different types of meridional circulation by choosing $\psi$ suitably.
For example, the one-cell meridional circulation used in many of the recent works
from our group (Karak 2010) is obtained by taking
\begin{eqnarray}
\label{eq:psi}
\psi r \sin \theta = \psi_0 (r - R_p) \sin \left[ \frac{\pi (r - R_p)}{(R_\odot -R_p)} \right]\{ 1 - e^{- \beta_1 \theta^{\epsilon}}\}\nonumber \\
\times\{1 - e^{\beta_2 (\theta - \pi/2)} \} e^{-((r -r_0)/\Gamma)^2} ~~~~
\end{eqnarray}\\
with
$\beta_1 = 1.5, \beta_2 = 1.3$, $\epsilon = 2.0000001$, $r_0 = (R_\odot - R_b)/3.5$, $\Gamma =
3.47 \times 10^{8}$ m, $R_p = 0.635 R_\odot$.
The value of $\psi_0/C$ determines the amplitude of the meridional circulation. On
taking $\psi_0/C=0.95{\times}15.0$, the poleward flow near the surface at mid-latitudes peaks around $v_0=15.0$ m s$^{-1}$.
The cell of the meridional circulation is confined between $R_p$ and $R_\odot$.  By making $R_p$ larger (but less
than $\Rs$), we can make the meridional circulation confined in the upper layers of the convection zone.

In order to have $N$ cells of meridional circulation, we can take a stream function of the form
\begin{equation}
\psi = \psi_1 + \psi_2 + \ldots + \psi_N,
\end{equation}
where each term in the stream function
gives rise to a cell of meridional circulation. Here we describe how we generate a two-cell
meridional circulation pattern, used in some of our simulations. The details of how we generate three-cell and
more complicated patterns are given in an Appendix. Since Zhao et al.\ (2013) claim that the upper cell of
the meridional circulation is confined above $0.82 R_{\odot}$, we do some calculations with two cells above
and below $R_m = 0.82 R_{\odot}$. To generate such a pattern of meridional circulation, we use the
stream function
\begin{equation}
{\psi}=\psi_u + \psi_l
\end{equation}\\
The stream function $\psi_u$ which generates the upper cell is given by
\begin{eqnarray}
\label{eq:psiu}
\psi_u = {\psi_{0u}}\left[1-{\rm erf}\left(\frac{r-0.91R_\odot}{1.0}\right)\right](r - R_{m,u})~~\nonumber \\
\times\sin \left[\frac{\pi (r - R_{m,u})}{(R_\odot -R_{m,u}}\right]\{ 1 - e^{- \beta_1 \theta^{\epsilon}}\}\nonumber \\
\times\{1 - e^{\beta_2 (\theta - \pi/2)} \} e^{-((r -r_0)/\Gamma)^2} 
\end{eqnarray}\\
where the parameters have the following values:
$\beta_1 = 3.5, \beta_2 = 3.3$, $r_0 = (R_\odot - R_b)/3.5$, $\Gamma =3.4 \times 10^{8}$ m, 
$R_{m,u} = 0.815 R_\odot$. The stream function $\psi_l$ which generates the lower cell is given by
\begin{eqnarray}
\label{eq:psil}
\psi_l = {\psi_{0l}}\left[1-{\rm erf}\left(\frac{r-0.95R_{m,l}}{1.8}\right)\right](r - R_p)~~\nonumber \\
\times\sin \left[\frac{\pi (r - R_p)}{(R_{m,l} -R_p)} \right]\{ 1 - e^{- \beta_1 \theta^{\epsilon}}\}\nonumber \\
\times\{1 - e^{\beta_2 (\theta - \pi/2)} \} e^{-((r -r_0)/\Gamma)^2}
\end{eqnarray}\\
where the parameters have the following values:
$\beta_1 = 3.2, \beta_2 = 3.0$, $r_0 = (\Rs - R_b)/3.5$, $\Gamma =3.24 \times 10^{8}$ m, $R_p = 0.65 R_\odot$, $R_{m,l} = 0.825 R_\odot$.  
We choose $\psi_{0u}/C$ and $\psi_{0l}/C$ in such a way that the velocity amplitudes in the upper and lower 
cells are around $20.0$ m s$^{-1}$ and $4.0$ m s$^{-1}$ respectively.

The two-cell meridional circulation given by the above expressions is shown in the upper part of Figure~2. Figure~2(a) shows
the streamlines of flow, whereas Figure~2(b) shows how $v_{\theta}$ varies with $r$ at the mid-latitude. The vertical dashed
lines in Figure~2(b) indicate bottoms and tops of the two cells.  It may be noted that both the cells have anti-clockwise
flow patterns.  This means that the flows at the bottom of the upper cell and at the top of the lower cell (which are
adjacent to each other) are in opposite directions involving a jump in the value of $v_{\theta}$ from one cell to
the next, as seen in Figure~2(b). If we replace $\psi_l$ by $-\psi_l$, then we can avoid such a jump
in the value of $v_{\theta}$.  This flow pattern
is shown in the upper part of Figure~3.  However, in this case, the meridional circulation at the tachocline is in the
poleward direction.  We shall see in Section~3 that this case will not give solar-like solutions.  If we want the meridional
circulation to be poleward at the surface and equatorward at the tachocline, and additionally we want to avoid a jump in
$v_{\theta}$, then we need at least three cells stacked one above the other in the radial direction. The flows in the top
and bottom cells have to be counter-clockwise, whereas the flow in the middle cell has to be clockwise.  The Appendix
presents the steam function that would give such a flow, which is shown in the upper part of Figure~4.  The results
with all these flow patterns are presented in the next Section, whereas results with a more complicated flow will be
presented in Section~4.   When we discuss the effects of changing the turbulent diffusivity in Section~5, we
shall describe how the diffusivity will be changed. Similarly, in Section~6 where we discuss the effects of 
turbulent pumping, we shall explain how pumping is included
in the mathematical theory.

\section{Results with radially stacked cells}

Let us first consider the situation that the \mc\ has a return flow at the middle of the convection zone
and there are no flows underneath it.  We generate such a \mc\ by taking $\psi = \psi_u$ with $\psi_u$ given
by Equation (11). The upper part of Figure~1 shows the streamlines and the profile of $v_{\theta}$ as a function
of $r$ at the mid-latitude.  The middle part of Figure~1 is the `butterfly diagram', which is essentially
a time-latitude plot of $B$ at the bottom of the convection zone.  The bottom part shows the radial field
at the solar surface as function of time and latitude.  In the butterfly diagram, we find that the belt of
strong $B$ propagates poleward rather than equatorward at the low latitudes, although there is a slight
tendency of equatorward propagation at the high latitudes. This result can be easily understood from the
Parker--Yoshimura sign rule, which holds when there is no flow at the bottom of the convection zone. We
have $\alpha$ positive in the northern hemisphere.  Since $\pa \Omega/ \pa r$ is positive in the low latitudes
and negative in the high latitudes (see Figure~1 of Chatterjee et al.\ (2004) showing the differential rotation
we are using), the Parker--Yoshimura sign rule implies poleward propagation at the low latitudes and equatorward
propagation at the high latitudes. In this case, we are completely unable to reproduce the solar behavior.
It may be noted that Guerrero \& de Gouveia dal Pino (2008) obtained solar-like behavior with a meridional
circulation similar to what we have used by including equatorward latitudinal pumping at the bottom of the
convection zone.  However, there are some uncertainties about the nature of latitudinal pumping 
at the present time and results of different simulations
often do not match each other (Racine et al.\ 2011).

\begin{figure}[!h]
\centering
\includegraphics[width=0.55\textwidth]{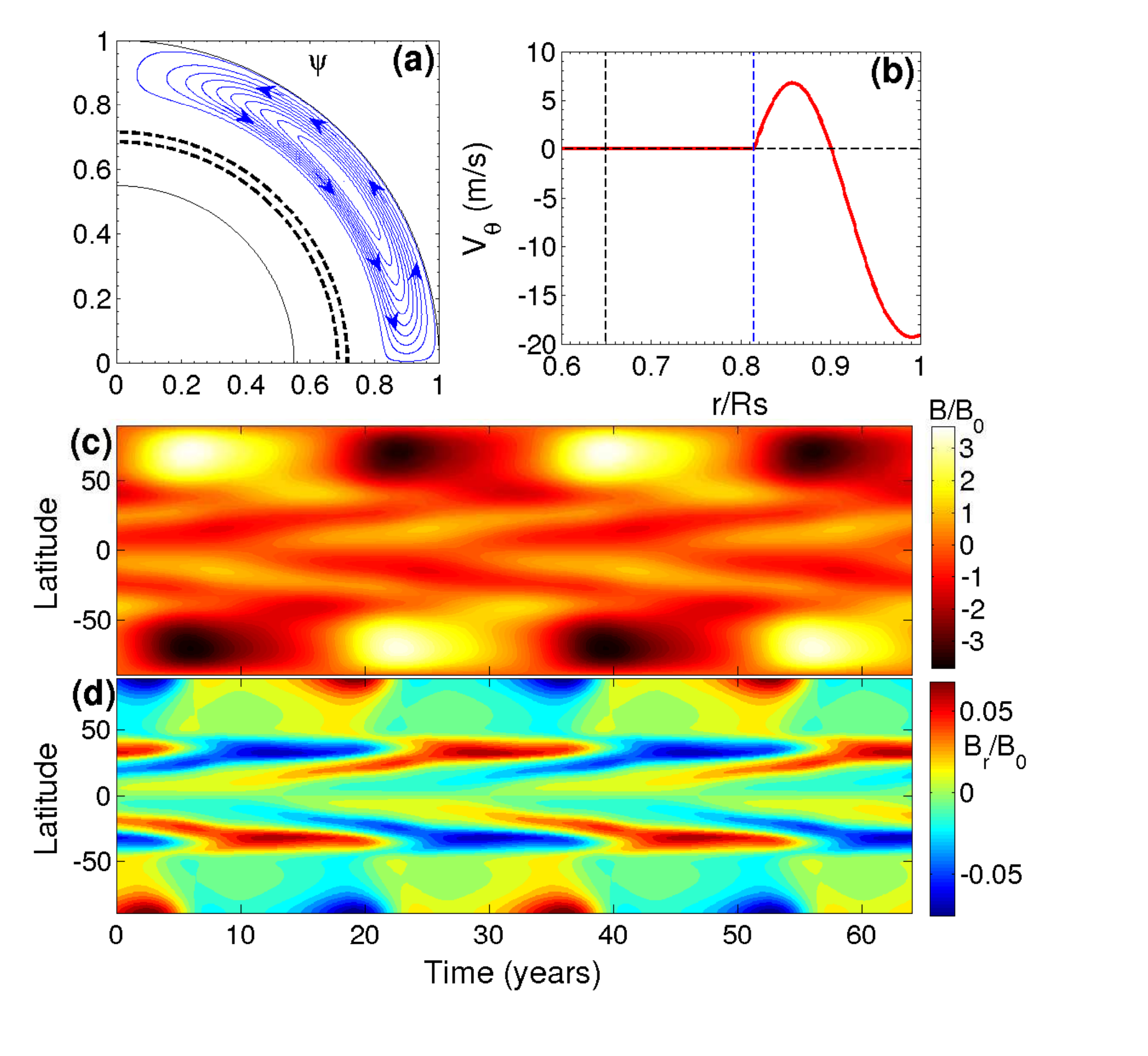}
\caption{(a) Streamlines of the shallow meridional circulation with no flow underneath. (b) $v_{\theta}$ as a function of ${r}/{\Rs}$ at the mid-latitude $\theta=45^{\circ}$.
(c) Butterfly diagram i.e, the time-latitude plot of the toroidal field at the bottom of the convection zone ($r=0.70\Rs$).
(d) Time-latitude plot of the radial field at the surface of the Sun. All the toroidal and radial fields are in the unit of $B_0$.}
\label{fig:rc2_nlwcll}
\end{figure}

\begin{figure}[!h]
\centering
\includegraphics[width=0.55\textwidth]{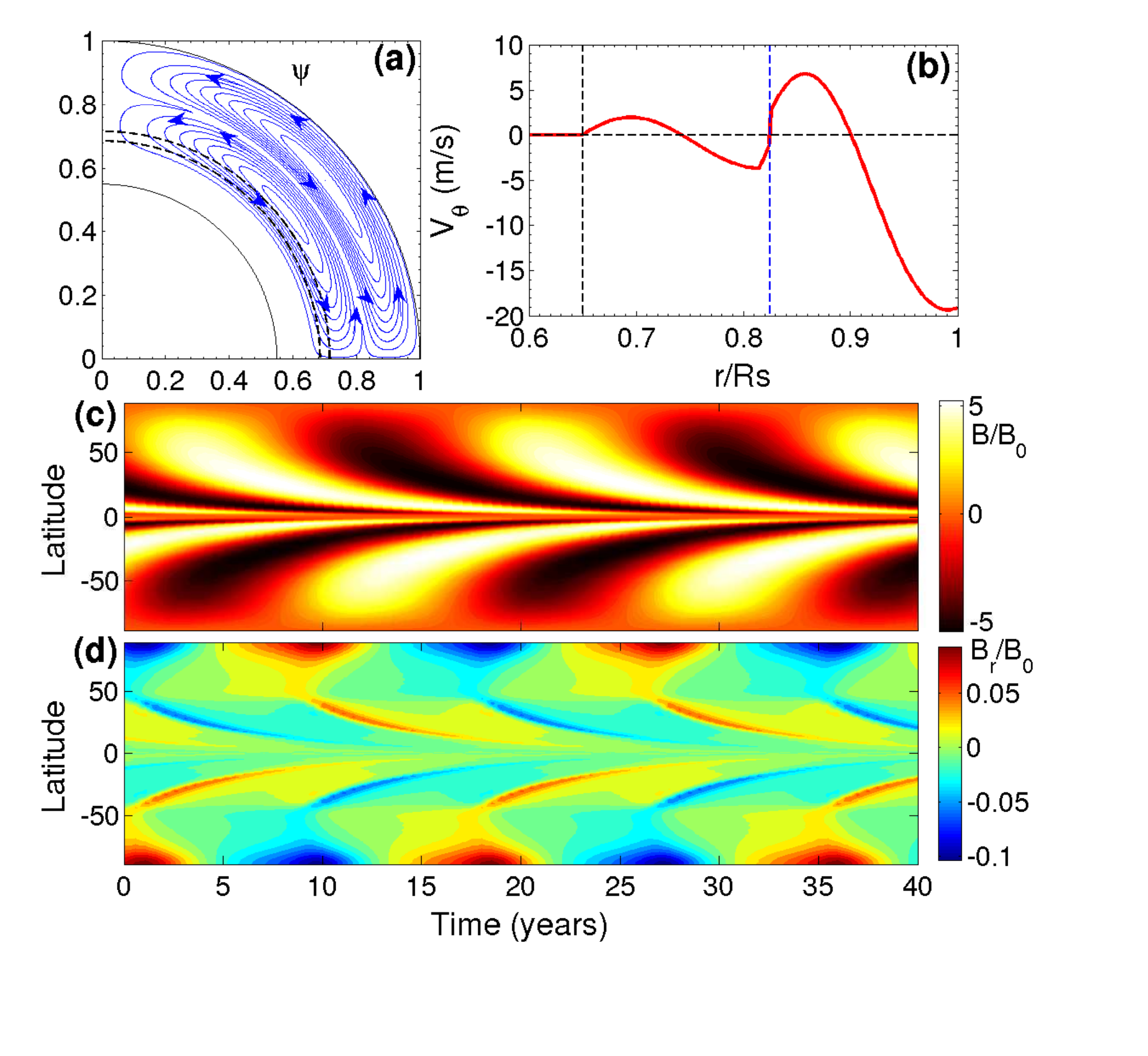}
\caption{(a) Streamlines for two radially stacked cells of \mc. Arrows show the direction of the flow. 
(b), (c) and (d) are the same plots as in Figure~1, for this \mc.}
\label{fig:radc2}
\end{figure}

Next we consider the two-cell pattern given by (10)--(12). For this case, Figure~\ref{fig:radc2} provides plots of the same things 
that are shown in Figure~\ref{fig:rc2_nlwcll} for the earlier case. We see that there is an equatorward flow at the bottom of the
convection  zone, although there is a jump in $v_{\theta}$ between the cells. We find that the equatorward flow at the
bottom forces an equatorward transport of $B$ in accordance with what we see in the Sun. Looking at the lowest
part of Figure~2, we also see that the polar field changes sign at the time of the sunspot maximum, in 
accordance with the observations.  Thus, on using the two-cell pattern with an equatorward flow at the
bottom of the convection zone, we can reproduce the equatorward migration of the sunspot zone as well as
the correct phase relationship between the toroidal and poloidal fields. It is true that the butterfly
diagram starts at a somewhat high latitude compared to what we see in the Sun.  It is well known that the
butterfly diagram can be confined more to lower latitudes by making the \mc\ a more penetrating
(Nandy \& Choudhuri 2002) and playing with other parameters. We had not bothered to fine-tune the 
parameters to achieve this, since our
main aim in this paper is to study the qualitative behavior of the system under various kinds of \mc.
Note from Figures~2(c)--(d) that the maximum $B$ at the bottom of the convection zone and
the maximum $B_r$ at the surface bear a ratio of about 100. It should be emphasized that this
ratio corresponds to smoothed mean field values of $B$ and $B_r$, which can have very different
values inside flux concentrations (Choudhuri 2003). 
 
We can avoid the jump in the value of $v_{\theta}$ seen in Figure~2(b)
by using a two-cell \mc\ in which $\psi_l$ is replaced by $-\psi_l$.
Since the flow at the bottom of the convection zone is poleward in this case, a study of this case
also throws light on the behavior of the dynamo with such a flow.  Our results are shown in Figure~3.
The butterfly diagram indicates poleward migration and the solar behavior is not reproduced in this
case.  The two-cell meridional circulation we have used is very similar to what was used
by Jouve \& Brun (2007) in one of their cases (see their Figure~2 and Figure~3).  The butterfly
diagram we have got is quite similar to what they got.

If we want to avoid a jump in $v_{\theta}$ and also to have a equatorward flow at the bottom
of the convection zone, then we need at least three cells of \mc\ stacked one over the other in the
radial direction.  The Appendix provides the mathematical prescription for generating such a \mc.
Figure~4 presents the results.  Since there is an equatorward flow at the bottom of the convection
zone, we again find that the solar behavior is reproduced, in the sense of having a butterfly diagram showing
equatorward migration as well as
the correct phase relationship between the toroidal and poloidal fields. 

\begin{figure}[!h]
\centering
\includegraphics[width=0.55\textwidth]{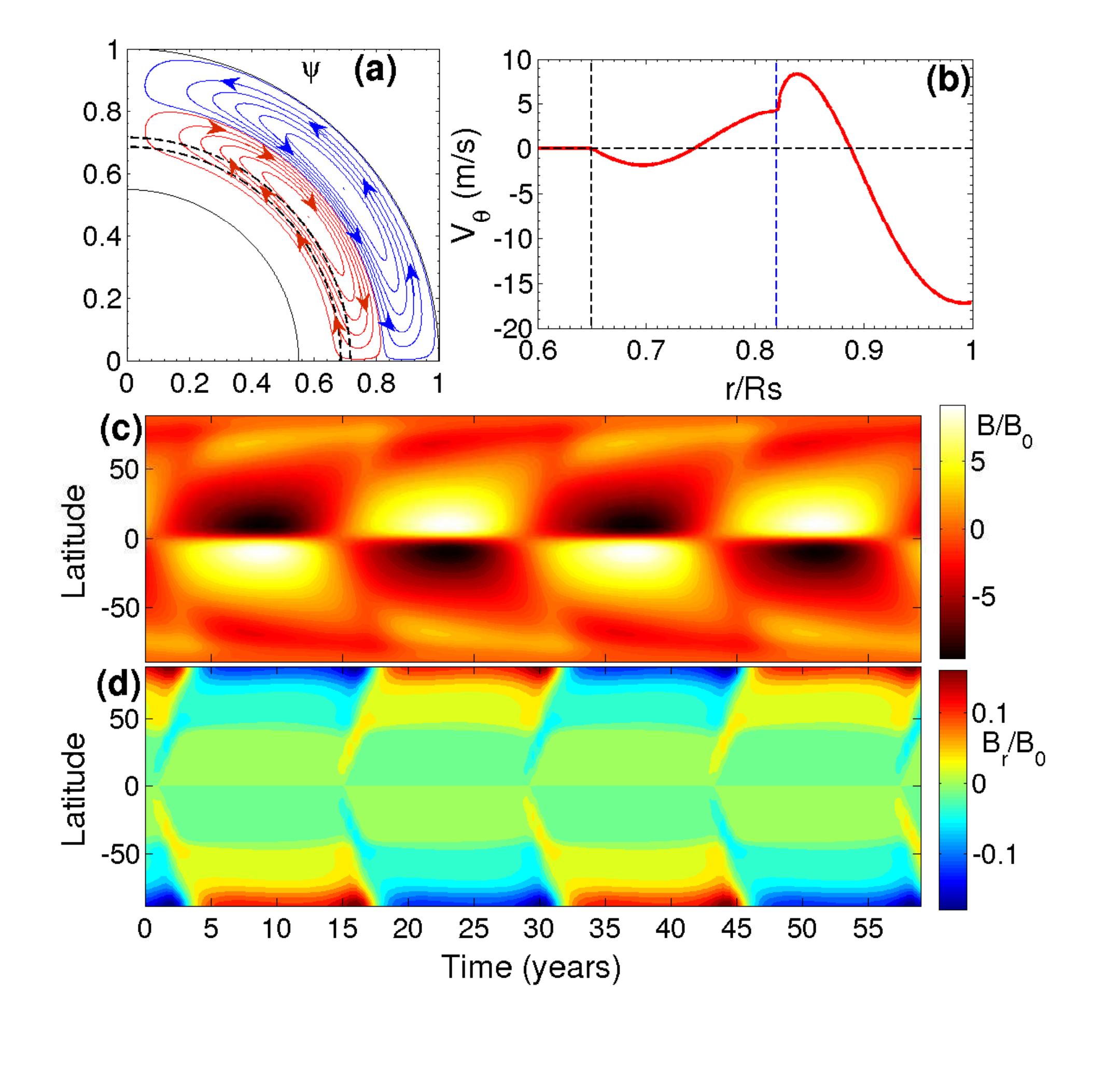}
\caption{(a) Streamlines for two radially stacked cells of \mc\ with circulations in the opposite
sense. Arrows show the direction of the flow. (b), (c) and (d) are the same plots as in Figure~1, for this \mc.}
\label{fig:radc2_rev}
\end{figure}

\begin{figure}[!h]
\centering
\includegraphics[width=0.55\textwidth]{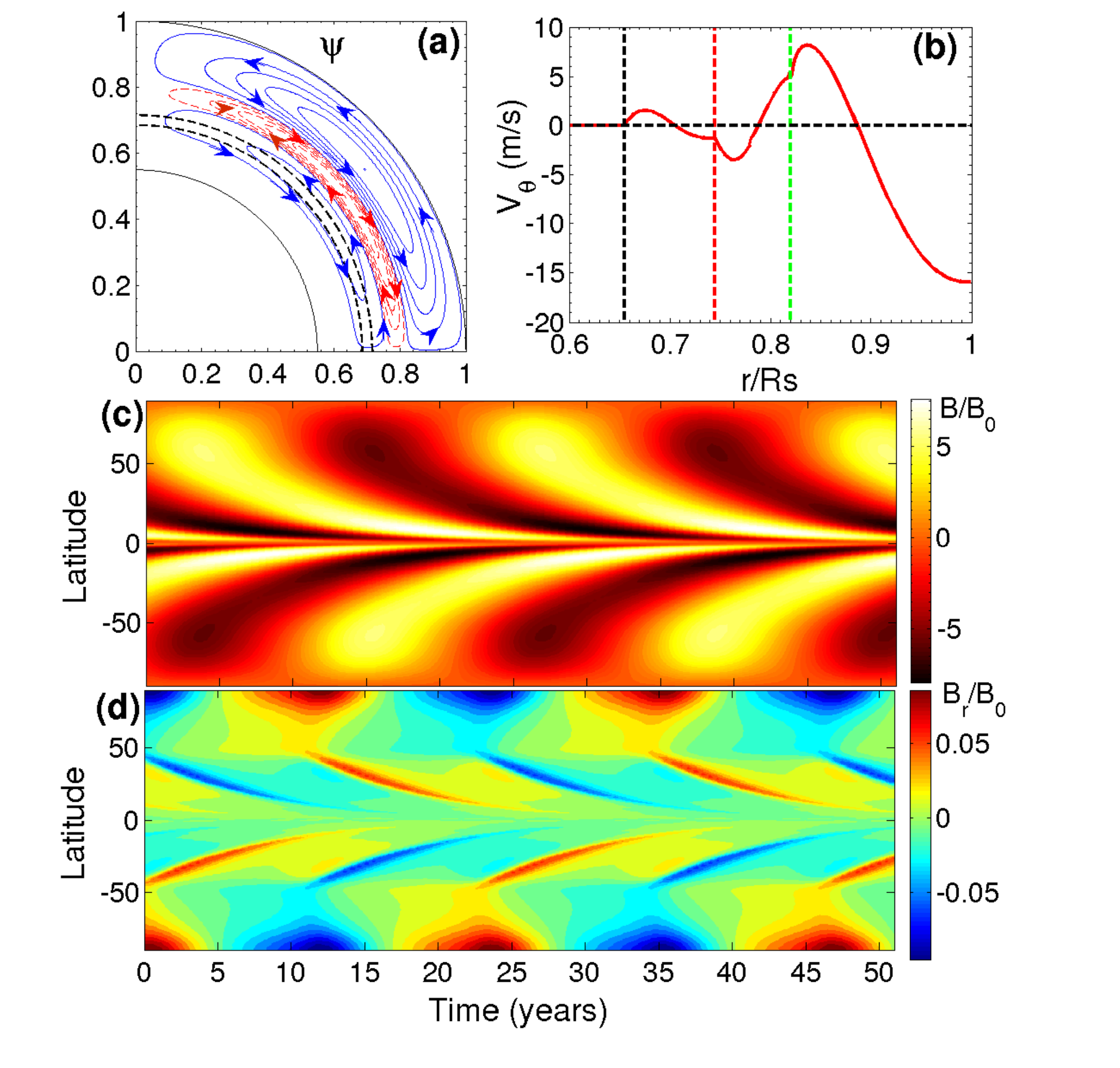}
\caption{(a) Streamlines for three radially stacked cells of \mc. Directions are shown by arrows. 
(b), (c) and (d) are the same plots as in Figure~1, for this \mc.}
\label{fig:radc3}
\end{figure}

One of the important results for the flux transport dynamo with a single cell of \mc\ is that
the period of the dynamo decreases when the \mc\ is made faster (Dikpati \& Charbonneau 1999; Karak 2010).
To explore how the dynamo period depends on the flow velocity in the multi-cell situation, we have
carried out a study for the case of three cells presented in Figure~\ref{fig:radc3}. We have carried out numerical
experiments by varying the flow amplitude of one cell, while keeping the flows in the other two
cells constant.  Figure~5 shows how the dynamo period changes with the change of the flow speed in
each of the three cells. It is clearly seen that the flow speeds in the upper two cells have very
minimal effect on the dynamo period.  It is the flow speed in the lowermost cell which determines
the dynamo period and we find the period to decrease with the increase in this flow speed ($T \sim v_{0,l}^{-0.72}$).  This
result can be easily understood from common sense, since the flow in the lowermost cell causes
the equatorward migration of $B$ (giving solar-like butterfly diagram), and it is no wonder that
the period becomes shorter on making this flow faster. 

\begin{figure}[!h]
\centering
\includegraphics[width=0.450\textwidth]{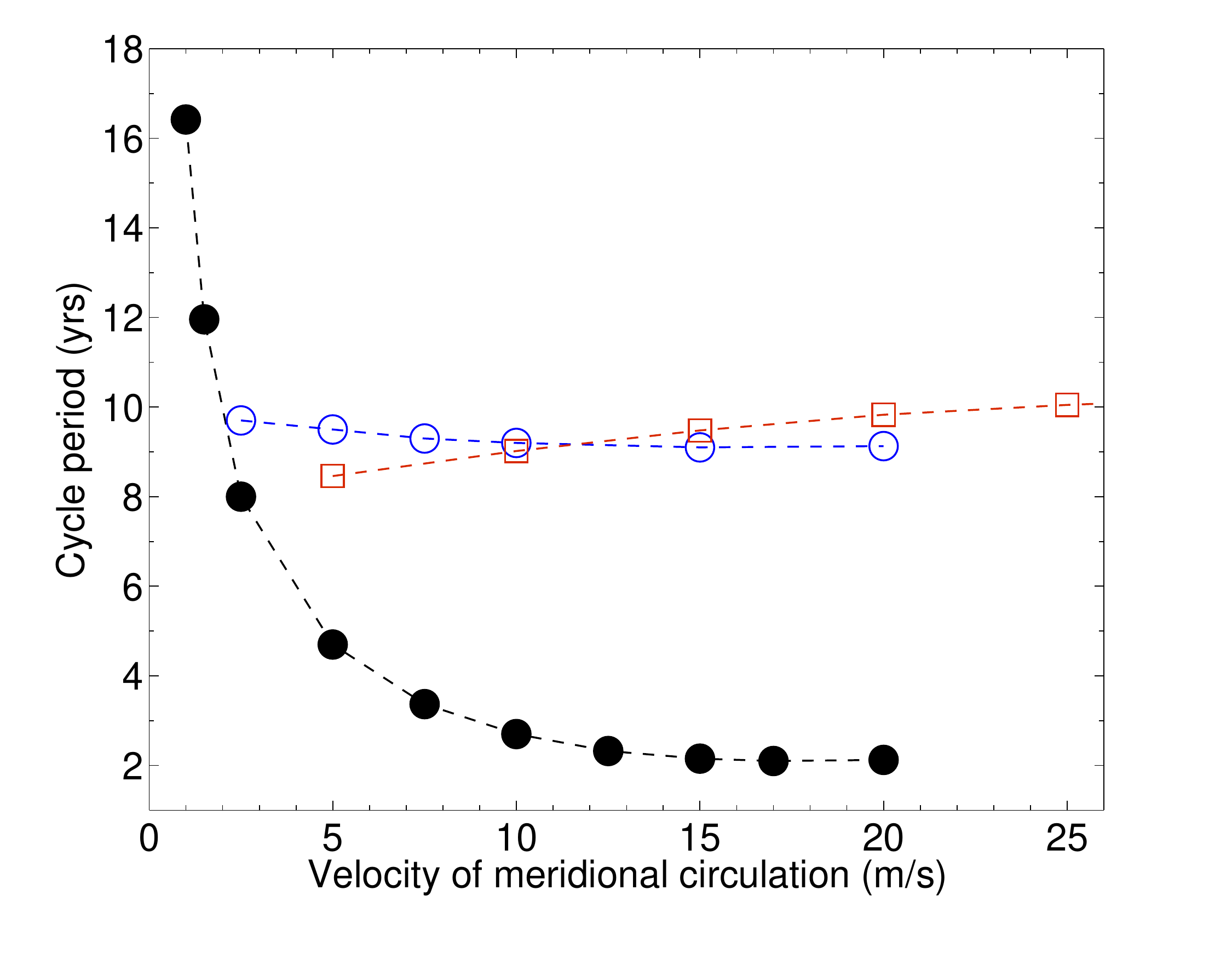}
\caption{Variation of solar cycle period with the velocity amplitudes of the three different cells shown in 
Figure~4(a). Black filled circles show the variation of the cycle
period with velocity amplitude of the lower cell while keeping velocities of other cells constant. 
Similarly blue circles show the variation of period with velocity amplitude of middle cell and red boxes for upper cell.  }
\label{fig:perio_vel}
\end{figure}

To sum up, as long as there is an equatorward flow at the bottom of the convection zone (the cases
of Figure~\ref{fig:radc2} and Figure~\ref{fig:radc3}), we are able to get solar-like behavior of the dynamo even if there
is a complicated multi-cell structure of the \mc, the period being determined by the flow
in the cell at the bottom of the convection zone.  Thus, even with a return flow of the \mc\ at a shallow
depth, the flux transport dynamo model can be made to work in this situation.
On the other hand, if there is no flow at
the bottom of the convection zone (the case of Figure~\ref{fig:rc2_nlwcll}) or if there is a poleward flow there
(the case of Figure~\ref{fig:radc2_rev}), then the dynamo model fails to reproduce solar behavior.  This conclusion
was obtained by considering multiple cells only in the radial direction. We consider more
complicated flows in the next Section and show that our main conclusion still holds.

\section{Results with more complicated cells}

We have carried out some simulations with fairly complicated multi-cell \mc, which reinforced
our main conclusion of the previous Section: we can have solar-like dynamo solutions as long
as there is an equatorward flow in low latitudes at the bottom of the convection zone. 
For the very complicated \mc\ pattern shown in Figure~6(a), we present the results in Figure~7. 
The Appendix indicates how this complicated flow is obtained from a suitable stream function.
Since there is an equatorward flow in low latitudes at the bottom of the convection zone,
we get solar-like butterfly diagrams even with such a flow.  It may be noted that the lowermost
cell in Figure~6(a) from the equator does not go all the way to the pole, but the cell ends at
some high latitude.  We find that this cell has to extend sufficiently to reasonably
high latitudes in order to give a solar-like butterfly diagram.  If the cell does not extend beyond
mid-latitudes, then we are unable to get very solar-like butterfly diagrams.
In Figure~6(b), we show a \mc\ with the lower cell not extending to high latitudes.  
Results for this case are presented in Figure~8. We see that the butterfly diagram
is much less realistic compared to the butterfly diagram presented in Figure~7.
It is clear from Figure~\ref{fig:most_comp} and Figure~\ref{fig:mcomp_2lowcl}
that the requirement for a solar-like butterfly diagram is that there has to be an equatorward
flow at the bottom of the convection zone having a sufficient latitudinal extent from the equator
to a reasonably high latitude.

\begin{figure}[!h]
\centering
\includegraphics[width=0.5\textwidth]{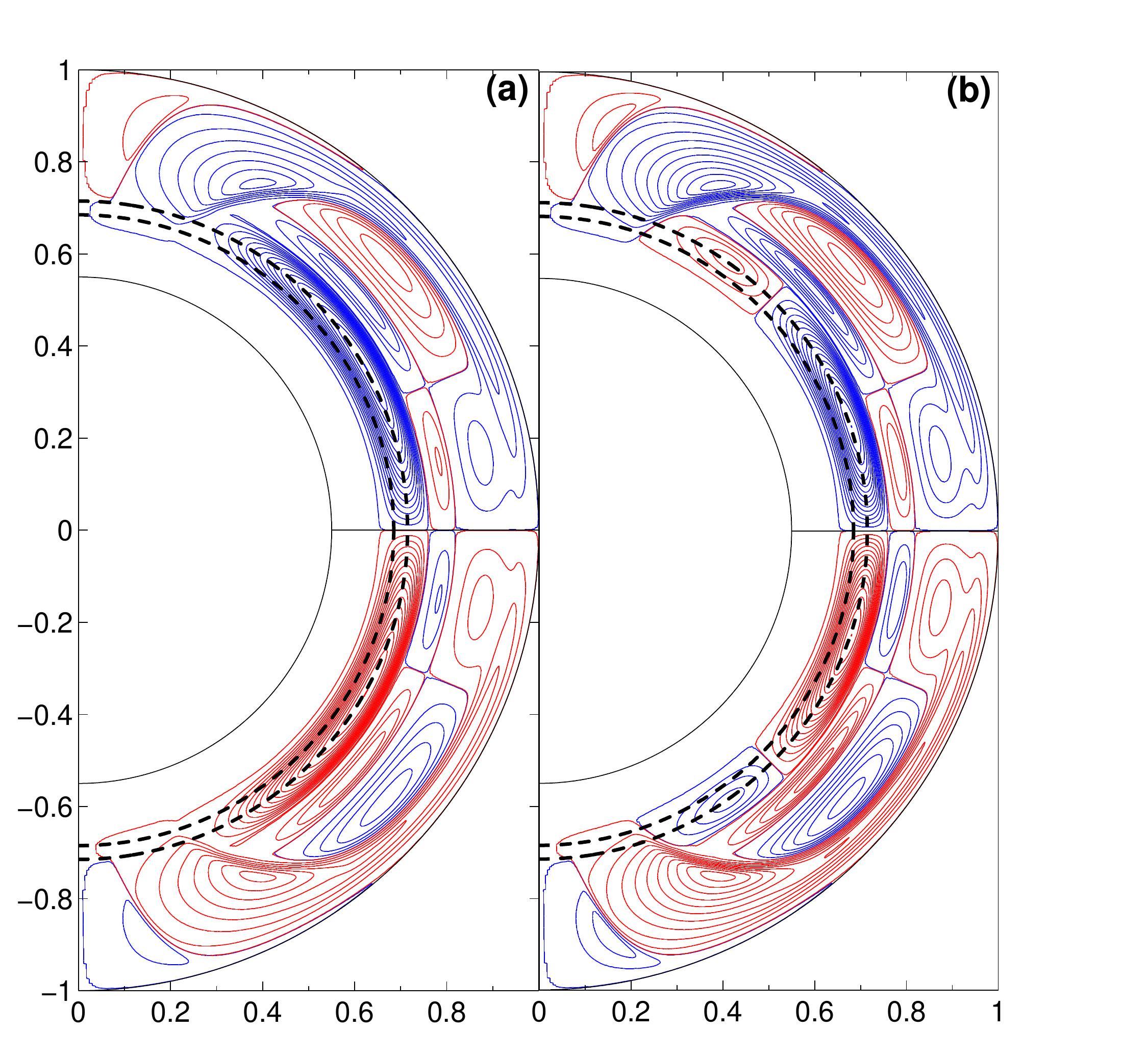}
\caption{Streamlines for two complicated \mc\ patterns.  The blue contours imply anti-clockwise
circulation, whereas the red contours imply clockwise circulation. The lowest cell in (a) extends
from the equator to fairly high latitudes, whereas this cell in (b) extends only to mid-latitudes. }
\label{fig:mcomp_cells}
\end{figure}

\begin{figure}[!h]
\centering
\includegraphics[width=0.50\textwidth]{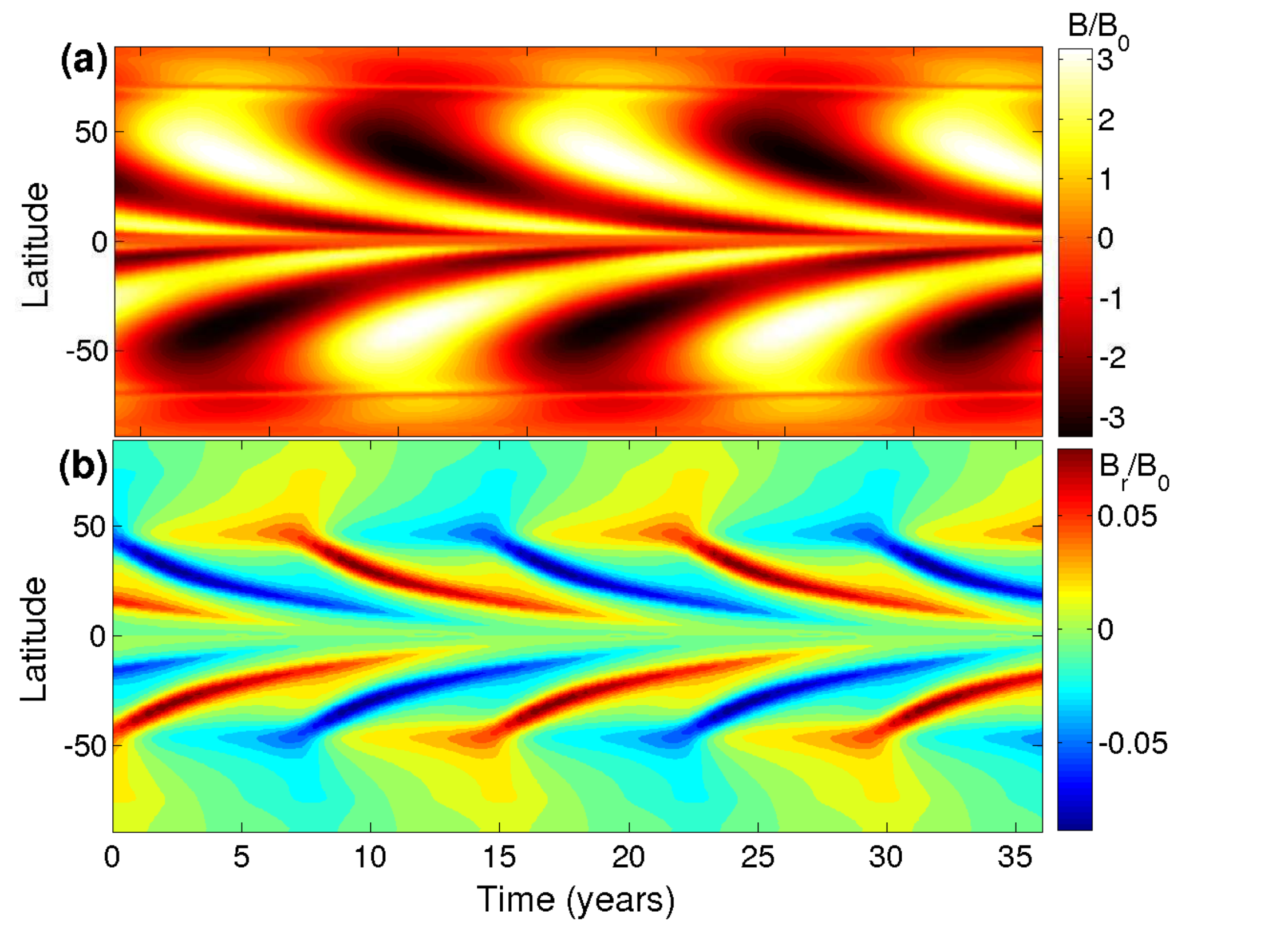}
\caption{(a) and (b) are the same plots as (c) and (d) in Figure~1, for the \mc\ given in
Figure~6(a).}
\label{fig:most_comp}
\end{figure}

\begin{figure}[!h]
\centering
\includegraphics[width=0.50\textwidth]{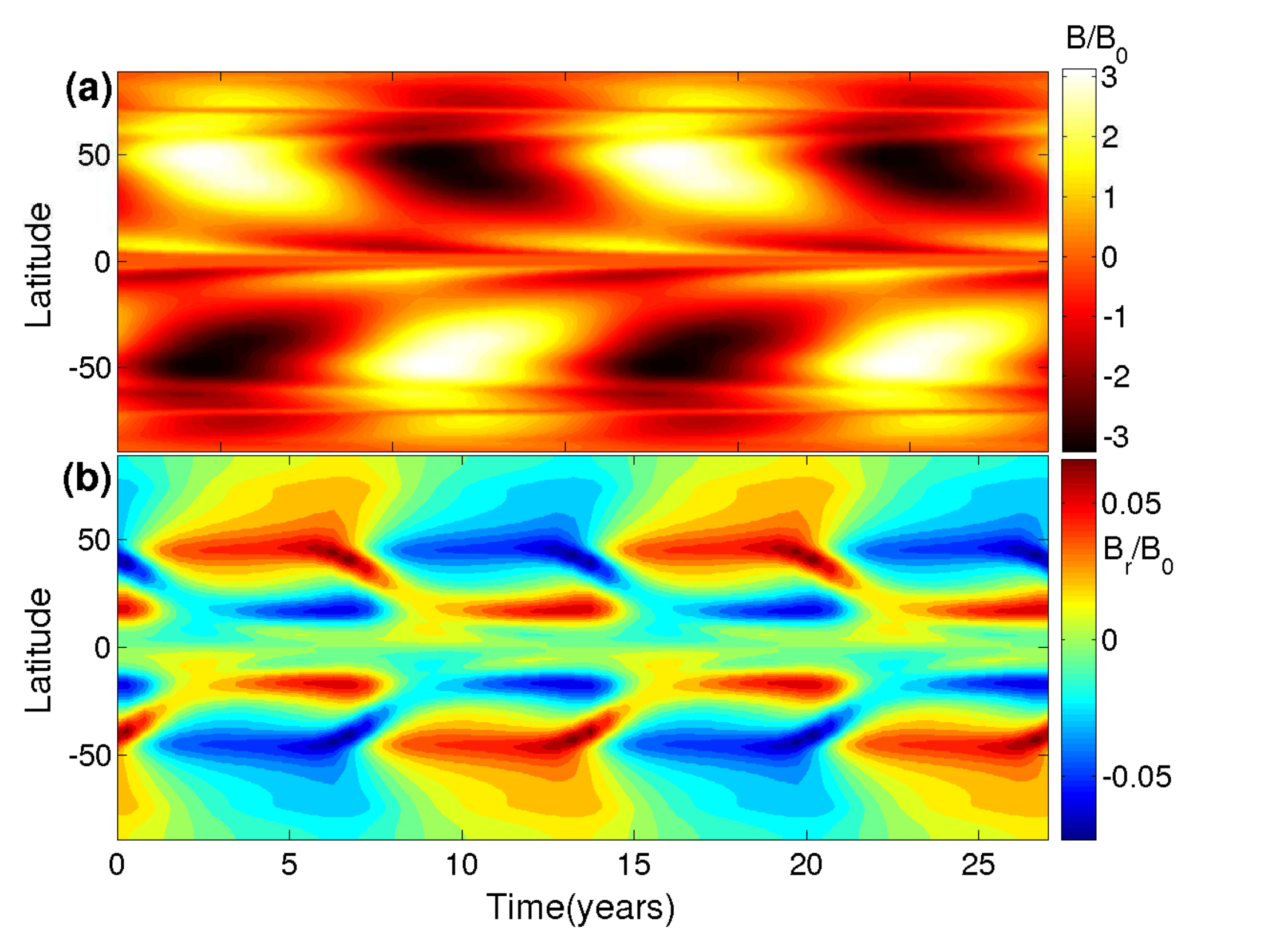}
\caption{Same as Figure~7, for the \mc\ given in Figure~6(b).}
\label{fig:mcomp_2lowcl}
\end{figure}

\section{Results for low diffusivity versus high diffusivity}

We have pointed out that the nature of the dynamo depends quite a bit on whether
the turbulent diffusivity within the convection zone is assumed to be high or low
(Jiang et al.\ 2007; Yeates et al.\ 2008; Hotta \& Yokoyama 2010; Karak 2010;
Karak \& Choudhuri 2011).  So far all the calculations in this paper have been
carried out with a diffusivity on the higher side.  With such diffusivity, the
poloidal field generated near the surface by the Babcock--Leighton mechanism 
reaches the bottom of the convection zone primarily due to diffusion and this
process is not affected by the presence of multiple cells. However, when the
diffusivity is low, it is the \mc\ which has to transport the poloidal field
from the surface to the bottom of the convection zone and such transport becomes
more complicated when there are multiple cells.
Now we come to the question
whether our main conclusion in the previous two sections holds when the diffusivity
is low.  Following Chatterjee et al.\ (2004), we specify the diffusivity 
for the high diffusivity case in the
following way: 
\begin{equation}
\eta_{p}(r) = \eta_{RZ} + \frac{\eta_{SCZ}}{2}\left[1 + \er \left(\frac{r - 0.7\Rs}
{0.03\Rs}\right) \right]
\label{eq:etap}
\end{equation}
\begin{eqnarray}
\label{eq:etat}
\eta_{t}(r) = \eta_{RZ} + \frac{\eta_{SCZ1}}{2}\left[1 + \er \left(\frac{r - 0.725\Rs}
{0.03\Rs}\right) \right]\nonumber \\
+ \frac{\eta_{SCZ}}{2}\left[ 1 + \er \left(\frac{r-0.975\Rs}{0.03\Rs}
\right) \right]
\end{eqnarray} 
Here $\eta_{RZ}$ is the diffusivity below the bottom of the convection zone which is
assumed to be small, whereas $\eta_{SCZ}$ and $\eta_{SCZ1}$ are respectively the
diffusivities of the poloidal and the toroidal components within the body of the
convection zone. Since the toroidal magnetic field is  believed to be much stronger than
the poloidal magnetic field, the diffusivity $\eta_{SCZ1}$ of the toroidal field is
assumed to be less than the diffusivity $\eta_{SCZ}$ of the poloidal field.
For high diffusivity case (all the results presented in \S~3 and \S~4), the values of the parameters for $\eta_{p}$ are  
$\eta_{RZ}=2.2 \times 10^8$ cm$^2$ s$^{-1}$, $\eta_{SCZ}=2.2\times10^{12}$ cm$^2$ s$^{-1}$,
and for $\eta_{t}$ are $\eta_{SCZ1}=4.0\times10^{10}$ cm$^2$ s$^{-1}$. Figure~9 shows these diffusivities as
functions of $r$, which have been used in the calculations of \S~3 and
\S~4. Now our aim in this Section is to study the case when the diffusivity
of the poloidal field is less.  To achieve this, we now take both $\eta_p$ and $\eta_t$
to be equal to $\eta_t$ in the high diffusivity case, as given by (14).  This means that
the diffusivity of the poloidal field within the main body of the convection zone is
now reduced by a factor of more than 50 (from $2.2\times10^{12}$ cm$^2$ s$^{-1}$ to
$4.0\times10^{10}$ cm$^2$ s$^{-1}$) for the studies presented in this Section.

\begin{figure}[!h]
\centering
\includegraphics[width=0.50\textwidth]{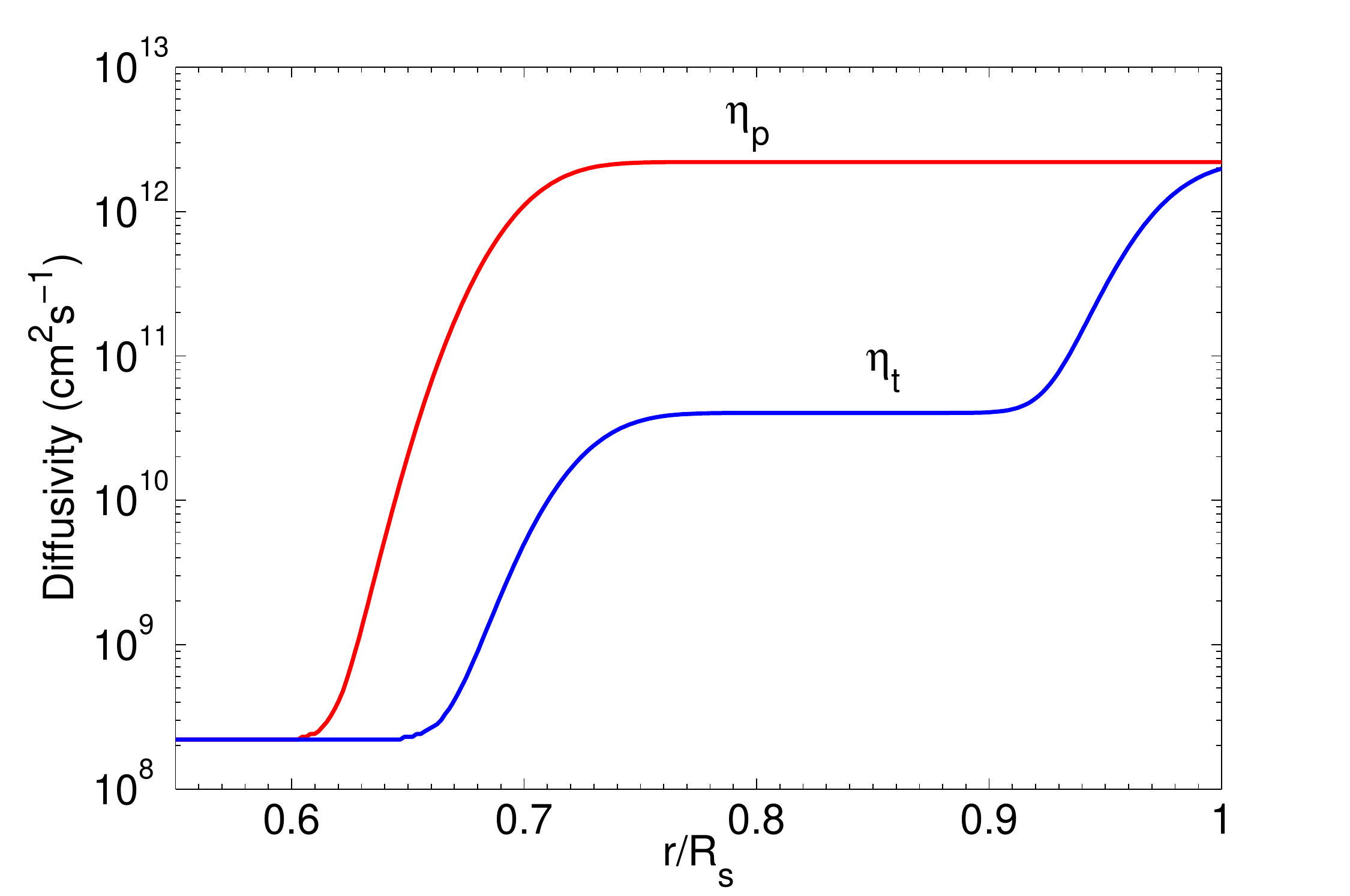}
\caption{Plots of $\eta_p(r)$ and $\eta_t(r)$ as given by (13) and (14). For the low diffusivity case, 
we take $\eta_{p}=\eta_{t}$.}
\label{fig:Highdif}
\end{figure}

% Since the toroidal component is  believed to be much stronger than
%the poloidal component, the diffusivity $\eta_{SCZ1}$ of the toroidal component is
%assumed to be less than the diffusivity $\eta_{SCZ}$ of the poloidal component, to
%account for the quenching of diffusivity when the magnetic field is strong. Table~1
%gives the values of $\eta_{RZ}$, $\eta_{SCZ}$ and $\eta_{SCZ1}$ for the two cases
%of high diffusivity (all the results presented in \S~3 and \S~4) and low diffusivity
%(some of the results in this Section) used in the present paper.

\begin{figure}[!h]
\centering
\includegraphics[width=0.50\textwidth]{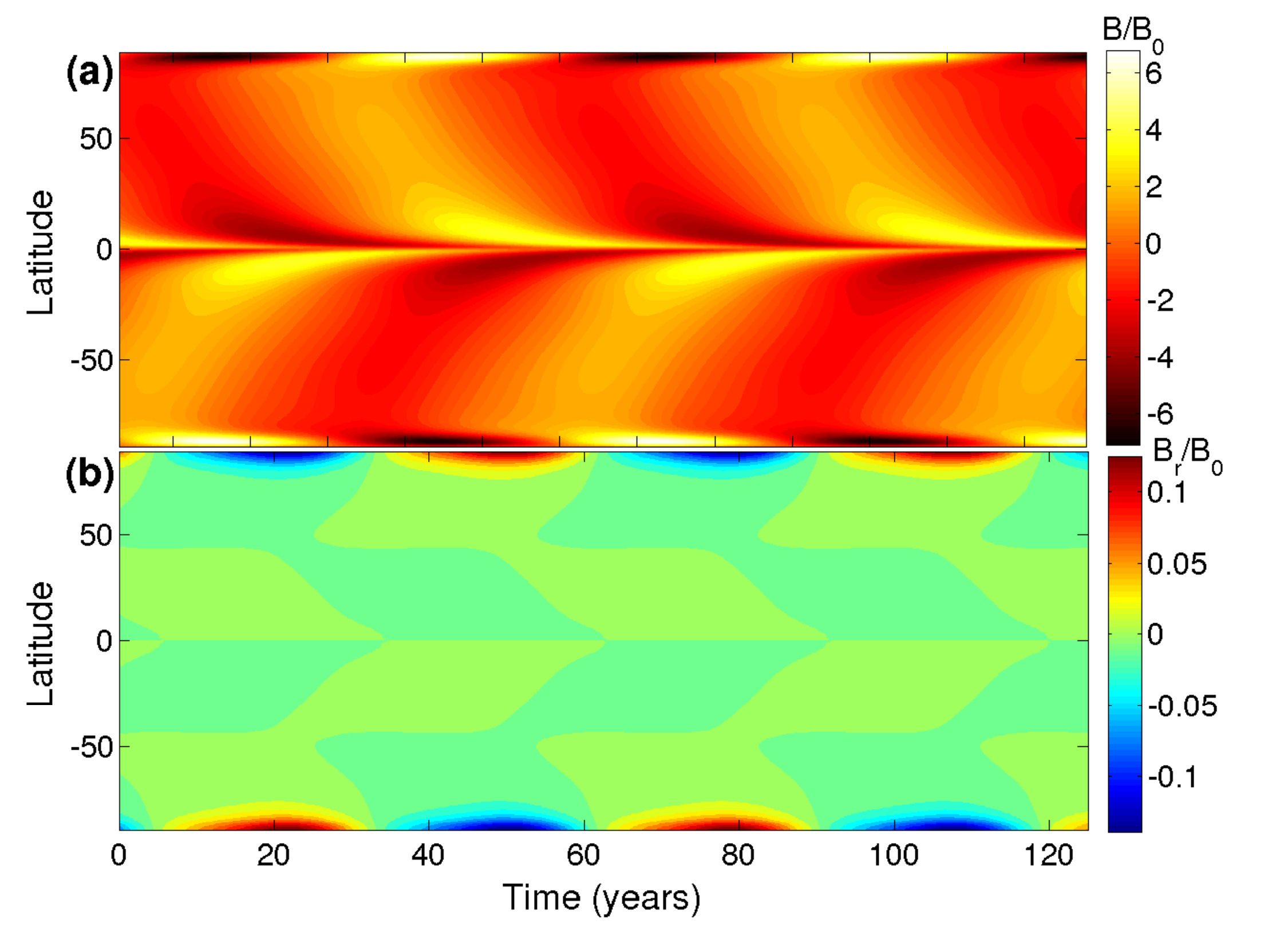}
\caption{Same as Figure~7, for the case of three radially stacked cells used in Figure~4
except that the diffusivity of the poloidal field is now lowered by making $\eta_p=\eta_t$.}
\label{fig:rc3lowdif}
\end{figure}

\begin{figure}[!h]
\centering
\includegraphics[width=0.70\textwidth]{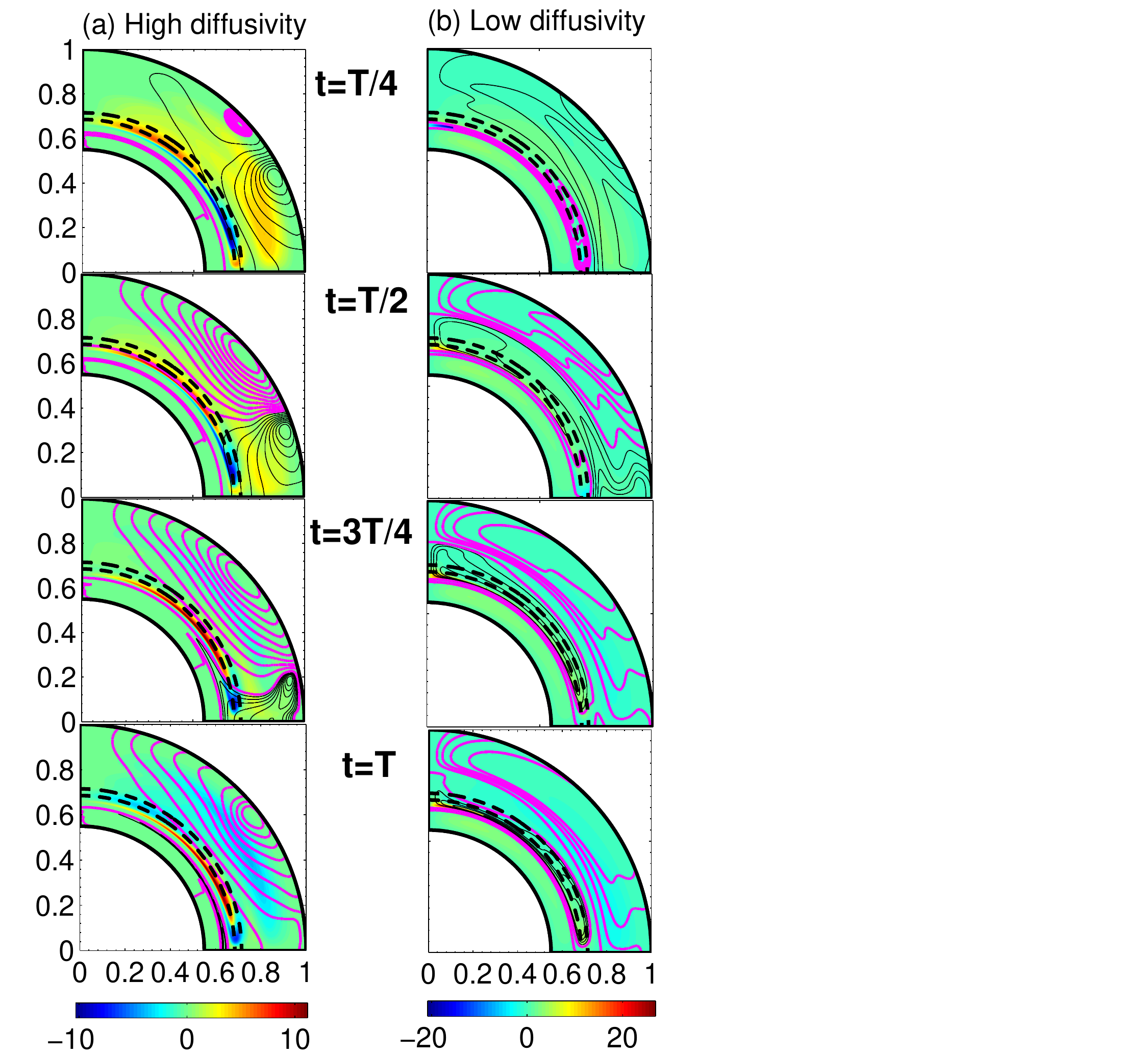}
\caption{The poloidal field lines at four different stages of a solar cycle for the cases
of (a) high diffusivity and (b) low diffusivity. The magenta and the black 
colors respectively indicate the clockwise and anti-clockwise sense of field lines.
The background colors indicate the strength of the toroidal field .}
\label{fig:pol_tor}
\end{figure}

To understand the effect of lowering the diffusivity, we carry on calculations
for the case of three radially stacked cells (the case shown in Figure~\ref{fig:radc3}) by
changing the diffusivity from the higher value to the lower value as mentioned
above. While reducing the diffusivity, we also reduce the strength of the
$\alpha$-coefficient as pointed out in \S~2. All the other parameters are kept unchanged. Figure~10 presents the
results. Although we still find solar-like butterfly diagrams, we find that
the period has become much larger on reducing the diffusivity.  This is not
surprising.  When the diffusivity is low, the poloidal field generated by the
Babcock--Leighton mechanism near the surface is transported to the bottom of
the convection zone (where the toroidal field is generated from it) by the \mc.
If there is only one cell, then this is easily accomplished.  However, when
there are three radially stacked cells as we are considering, the situation
becomes much more complicated.  The uppermost cell brings the poloidal field
from the surface to its bottom. From there, the middle cell has to advect the
poloidal field to its bottom.  Finally, the lowermost cell takes the poloidal
field to the bottom of the convection zone. In this process, the period of the
dynamo gets lengthened.  Figure~\ref{fig:pol_tor} shows how the poloidal field lines evolve
with the cycle for the case of three radially stacked cells---both when the
diffusivity is high (the case of Figure~\ref{fig:radc3}) and when the diffusivity is
low (the case of Figure~\ref{fig:rc3lowdif}). In the high diffusivity case, the poloidal field
generated at the surface is transported downwards to the bottom of the convection
zone by diffusion.  Hence, in this case, we find that the poloidal field lines
still are not very different from what we find in the case of \mc\ with one
cell, as shown in Figure~4 of Jiang et al.\ (2007).  However, when the 
diffusivity is low, the poloidal field is nearly frozen during a cycle
and is advected by the \mc. In a three-cell \mc, we find that the poloidal
field becomes very complicated, as seen in the right column of Figure~\ref{fig:pol_tor}.

It has been pointed out that, when we introduce fluctuations to model
irregularities of solar cycle, the dynamo models with high and low diffusivities
behave completely differently (Jiang et al.\ 2007; Karak \& Choudhuri 2011). In the high diffusivity
model, the fluctuations diffuse all over the convection zone in time scale
comparable to the period of the dynamo.  On the other hand, fluctuations
in the low diffusivity model remain frozen during the period of the
dynamo.  Jiang et al.\ (2007) explained how the observed correlation between the
polar field during a sunspot minimum and the strength of the next cycle arises
in the high diffusivity model.
This correlation, which forms the basis of solar cycle prediction in the high
diffusivity model, does not exist in the low diffusivity model.  We now
check if these results hold even when we have multiple cells of the \mc.
Choudhuri et al.\ (2007) identified the fluctuations in the Babcock--Leighton
process as the main source of irregularity in the sunspot cycles. These fluctuations
arise from the scatter in the tilt angles of sunspots caused by the effect
of convective turbulence on rising flux tubes (Longcope \& Choudhuri 2002). To model
these fluctuations, we introduce stochastic fluctuations in $\alpha_0$ appearing in (5). 
We set 
\begin{equation}
\alpha_0 \equiv \overline{\alpha_0} [1 \pm 0.75  \sigma (\tau_{\rm{cor}})],
\end{equation} 
where $\sigma$ 
is a uniformly generated random number within 0 to 1 which changes value after a coherence time $\tau_{cor} = 1$ month.
This makes $\alpha_0$ to fluctuate randomly around its mean value $\overline{\alpha_0}$ 
with 75\% amplitude of fluctuations. A simulation with such stochastic fluctuations in $\alpha$
in a traditional $\alpha \Omega$ dynamo model was first presented by Choudhuri (1992).

To study the correlation between the polar field and the strength of the next cycle, we 
consider the procedure of Yeates et al.\ (2008). We calculate the 
correlation between the peak of the surface radial flux $\phi_{r}$ at high latitudes of a cycle 
with that of the peak value of the deep-seated toroidal flux $\phi_{tor}$ of next cycle. We take 
$\phi_{r}$ as the flux of radial field over the solar surface from latitude $70^{\circ}$ to $89^{\circ}$, 
and $\phi_{tor}$ as the flux of toroidal field over the region $r=0.677\Rs$ -- $0.726\Rs$ and latitude $10^{\circ}$ to $45^{\circ}$.
In the case of a one-cell \mc\ 
(not presented in detail in this paper), we get a strong correlation between the high-latitude radial 
flux at the end of a cycle with the toroidal flux of the next cycle, with a correlation coefficient of 0.79, 
which is comparable to the result of Jiang et al.\ 2007 (see their Figure~5) and Yeates et al.\ (2008) (see their Fig.\ 11b).
Interestingly, for radially stacked three cells also, we get a strong correlation 
of 0.75 for the high diffusivity case.
Figure~12 shows this result, along with the result for the low 
diffusivity case. For the low diffusivity case, the correlation is substantially poorer.
Thus, a multi-cell \mc\ not only reproduces the regular periodic features of a
simple flux transport dynamo model, it also reproduces some of the irregular
features of the cycle if the diffusivity is high.  The methodology for
predicting the next cycle developed by Choudhuri et al.\ (2007) and Jiang
et al.\ (2007) should work approximately the same way 
in the high diffusivity model even when the \mc\ has
a complicated multi-cell structure.
\begin{figure}[!h]
\centering
\includegraphics[width=0.65\textwidth]{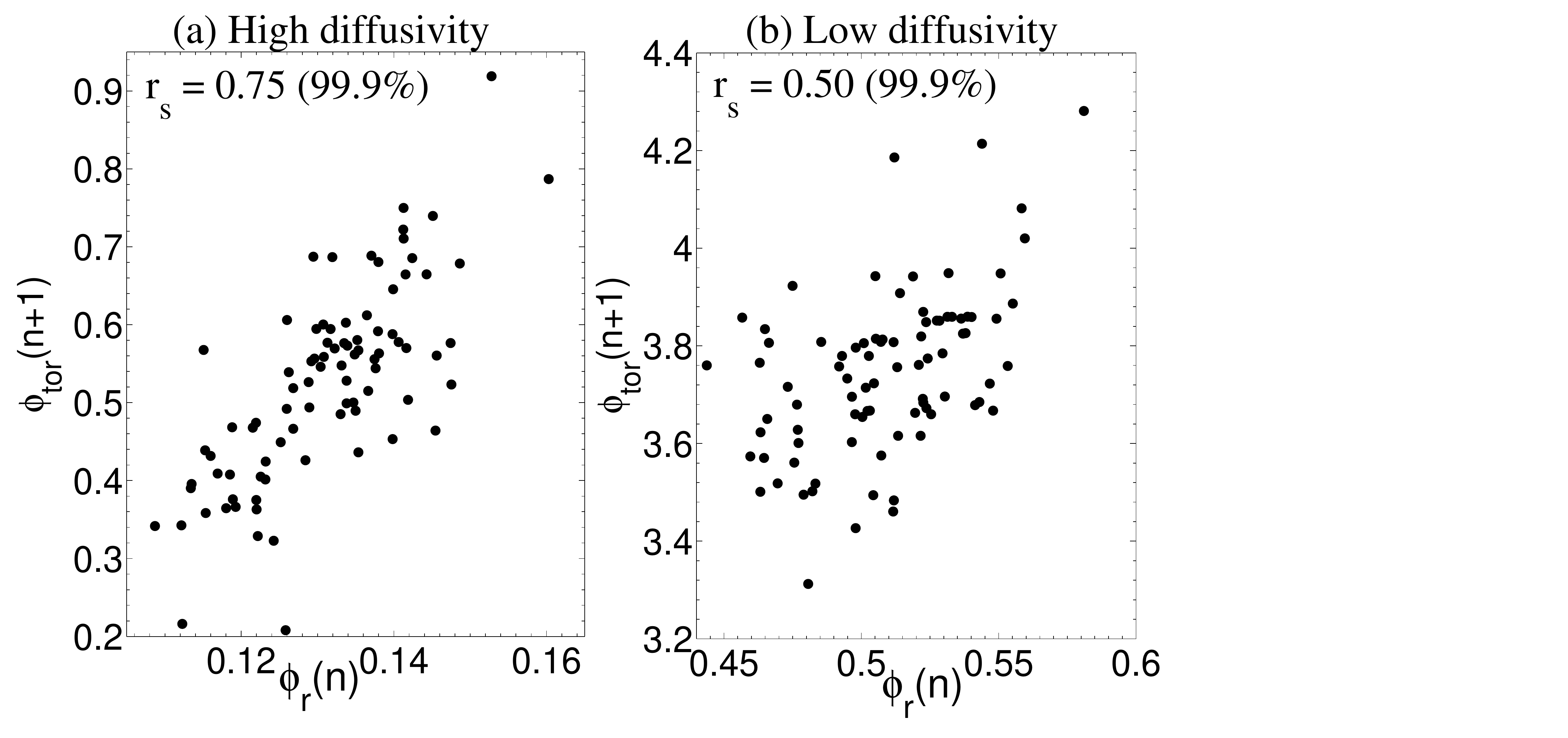}
\caption{Correlation between peak polar flux strength at the end of the $n$-th cycle and the peak toroidal flux strength of the 
$(n+1)$-th cycle for (a) high diffusivity and (b) low diffusivity cases. }
\label{fig:corr}
\end{figure}

\section{The effect of turbulent pumping}

\begin{figure}[!h]
\centering
\includegraphics[width=0.55\textwidth]{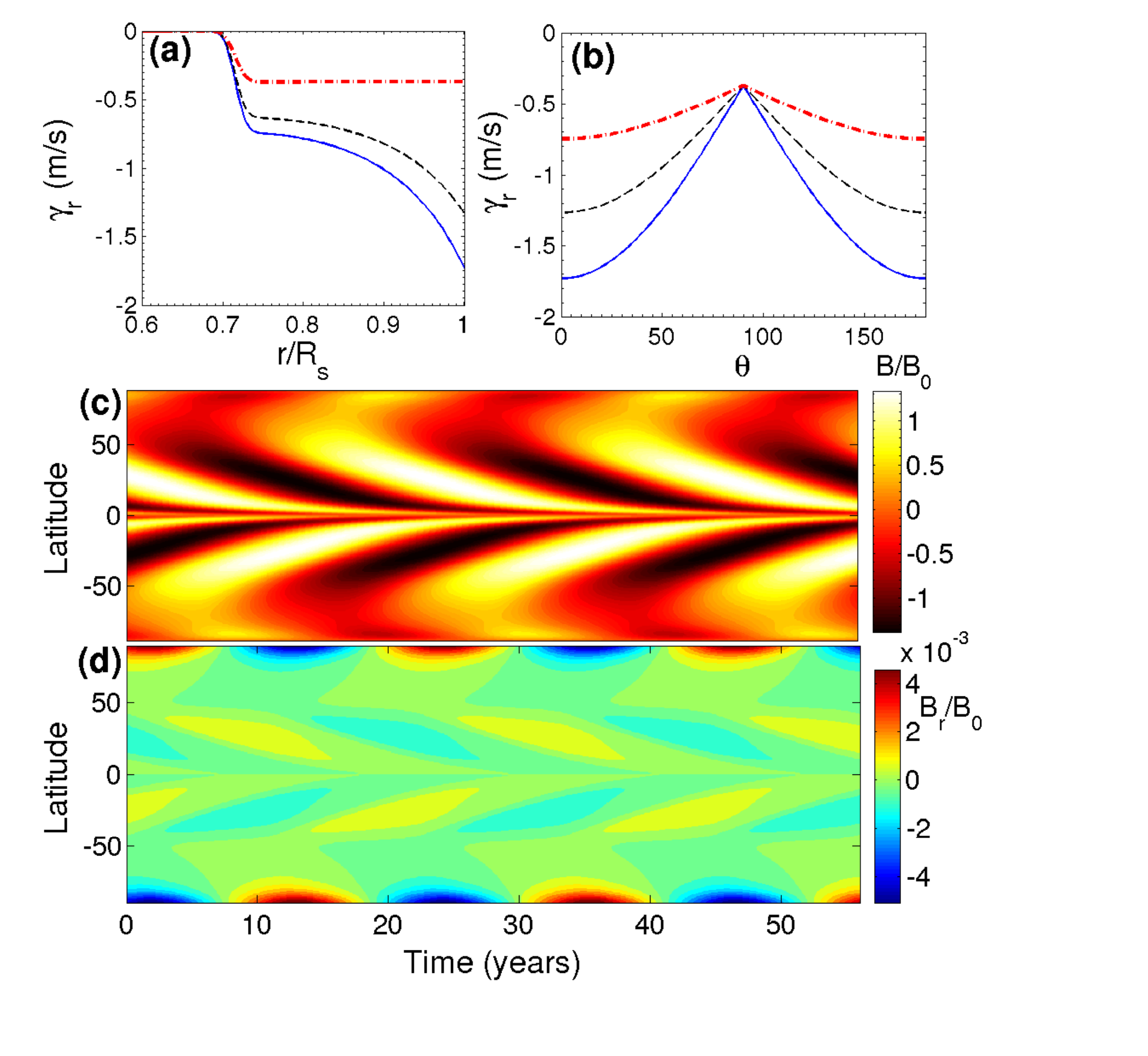}
\caption{(a) Radial pumping $\gamma_r$ as a function of radius at different co-latitudes ($\theta$). The solid (blue),
the dashed (black) and the dash-dotted (red) lines correspond to $\theta = 90^{\circ}, 45^{\circ}$ and $0^{\circ}$ respectively.
(b) $\gamma_r$ as a function of $\theta$ at different radii.  The solid (blue),
the dashed (black) and the dash-dotted (red) lines correspond to $r = \Rs, 0.95\Rs$ and $0.75\Rs$ respectively.
(c) and (d) are the same as (a) and (b) in Figure~10 with the radial pumping added now.}
\label{fig:pumping}
\end{figure}

One possible mechanism for transporting magnetic fields across the solar convection zone which
we have so far not included in our paper is turbulent pumping. Many theoretical as well as numerical studies
indicated that, in the strongly stratified solar convection zone, the magnetic fields can 
be pumped preferentially downward towards the base of the convection zone 
(Brandenburg et al.\ 1996; Tobias et al.\ 1998). 
Several magnetoconvection simulations have detected a downward pumping speed of a
few meters per second in the solar convection zone (Ossendrijver et al.\ 2002; K\"apyl\"a et al.\ 2006; Racine et al.\ 2011).
Guided by these studies, we now include the effect of turbulent pumping in our dynamo model
by introducing the following downward pumping velocity: 
\begin{eqnarray}
\gamma_r = - 0.1854 \left[ 1 + \rm{erf}\left( \frac{r - 0.715\Rs}{0.015\Rs}\right) \right]\nonumber \\ 
\times\left[ \rm{exp}\left( \frac{r-0.715\Rs}{0.25\Rs}\right) ^2 \rm{cos}\theta +1\right],
\label{rpumping}
\end{eqnarray}
the unit being m s$^{-1}$.
The variations of $\gamma_r$ as functions of radius and co-latitude are shown in 
the upper part of Figure~\ref{fig:pumping}.
Turbulent pumping appears as an advective term in the magnetic field equations. Therefore 
in (2) and (3) we add this extra term $\gamma_r$ in the radial velocity, i.e., we take $v_r \equiv v_r + \gamma_r $.
As in Karak \& Nandy (2012), Kitchatinov \& Olemskoy (2012) and Jiang et al.\ (2013), we 
first present results including only the radial pumping and not the latitudinal pumping.

Since the downward transport of the poloidal field by diffusion is reasonably efficient in
the high diffusivity model, the effect of downward turbulent pumping is not very pronounced
in this model.  However, in the low diffusivity model, the poloidal field is advected by
the \mc\ in the absence of turbulent pumping and the addition of downward pumping can have
quite dramatic effects.  Karak \& Nandy (2012) found that many of the differences between the
high and the low diffusivity models disappear on inclusion of downward turbulent pumping.
We have seen in \S~5 that the low diffusivity model with multi-cell \mc\ gives results which
do not match observations as closely as the results obtained with high diffusivity.
Although the case of \mc\ with three radially stacked cells even with low diffusivity
produces reasonably good equatorward propagation of toroidal field  at low latitude, the solar cycle
period becomes very long (see Figure~\ref{fig:rc3lowdif}).
We now repeat this calculation for the low diffusivity case by including the downward 
pumping.  The butterfly diagram is shown in the middle of Figure~13.  We find that the
period has become much shorter and the butterfly diagram looks quite similar to the butterfly
diagram of Figure~4 in the high diffusivity case.  Thus, even when multiple cells are
present in the \mc, the inclusion of downward turbulent pumping makes the results of 
the low diffusivity case quite similar to the results of the high diffusivity case.

A few magnetoconvection simulations (Ossendrijver et al.\ 2002; K\"apyl\"a et al.\ 2006; Racine et al.\ 2011) have detected a 
latitudinal turbulent pumping when rotation becomes important. However, while there is 
a general consensus about the radial downward pumping in the results of different groups,
the results of latitudinal 
turbulent pumping are more uncertain and its physical origin is not easy to understand. 
In view of these uncertainties, we do not carry out a detailed study of the effects of
latitudinal pumping in this paper.  As we already mentioned, Guerrero \& de Gouveia Dal Pino (2008)
found that they can get solar-like solutions on including a suitable equatorward latitudinal
pumping even when the return flow of the meridional circulation is at a shallow depth and
there is no flow below it.  We checked that we also get the same result when we include
the latitudinal pumping used by Guerrero \& de Gouveia Dal Pino (2008) in our model.\\

\section{Conclusion}
In the flux transport dynamo model which has been very successful in modeling
different aspects of the solar cycle, the \mc\ of the Sun is a crucial ingredient.
The major uncertainly in the flux transport dynamo model at the present time is
our lack of knowledge about the nature of the \mc\ in the deeper layers of the
convection zone. Although two-dimensional models can never treat magnetic buoyancy
and the Babcock--Leighton mechanism in a fully satisfactory, we believe that
these uncertainties are not so serious because different treatments of magnetic
buoyancy and the Babcock--Leighton mechanism give qualitatively similar
results (Nandy \& Choudhuri 2001; Choudhuri
et al.\ 2005). Since the \mc\ arises out of a delicate imbalance between the
centrifugal forcing and the thermal wind (Kitchatinov 2011), it is challenging to
model it theoretically.  Models of differential rotation based on a mean field
treatment of turbulence give rise to a \mc\ (Kitchatinov \& R\"udiger 1995; Rempel 2005;
Kitchatinov \& Olemskoy 2011b).  MHD simulations of convection with the dynamo
process also produce meridional circulations (Brown et al.\ 2010; Racine et al.\ 2011; Warnecke et al.\ 2013; K\"apyl\"a et al.\ 2013).
In such simulations, the \mc\ is often found to have several cells and to vary rapidly
with time.  We are still far from having a definitive theoretical model of the
Sun's \mc.

Most of the flux transport dynamo models are based on the assumption of a
single-cell \mc\ having a return flow at the bottom of the convection zone.
While a support for such a return flow may have been missing, this assumption
of a deeply penetrating single-cell \mc\ was at least consistent with all the
observational data available till about a couple of years ago. The equatorward
propagation of the sunspot belt was indeed regarded as indicative of the \mc\ flow
velocity at the bottom of the convection zone (Hathaway et al.\ 2003). Only
recently there are claims that the \mc\ may have a return flow at a much shallower
depth (Hathaway 2012; Zhao et al.\ 2013).  If these claims are corroborated by
independent investigations of other groups, then we shall have to conclude that
the assumption of a deep one-cell \mc\ is not correct.  Since this assumption
was extensively used in most of the kinematic flux transport dynamo models, a
crucial question we face  now is whether this assumption is so essential that
the flux transport dynamo models would not work without this assumption or whether the
flux transport dynamo models can 
still be made to work with a suitable modification of this
assumption. 

On the basis of our studies, we reach the conclusion that, in order to have a
flux transport dynamo giving a solar-like butterfly diagram, we need an equatorward
flow in low latitudes at the bottom of the convection zone. This flow is essential
to overcome the Parker--Yoshimura sign rule and to advect the toroidal field generated
in the tachocline in the equatorward direction.  As long as there is such a flow,
we find that the flux transport dynamo works even if the \mc\ has a much more
complicated structure than what has been assumed in the previous models.  If there
is a return flow at a shallow depth and there are no flows underneath, then the 
flux transport dynamo will not work. If there is a poleward flow at the bottom
of the convection zone, then also we do not get solar-like butterfly diagrams.
However, underneath a shallow return flow if we have multiple cells in such a 
way that there is an equatorward flow in low latitudes at the bottom of the 
convection zone, then the flux transport dynamo works without any serious problem.
The assumption of such a multi-cell \mc\ does not contradict any observational
data available at the present time.  MHD simulations also support the existence of a
complicated multi-cell \mc\ (Brown et al.\ 2010; Racine et al.\ 2011; Warnecke et al.\ 2013; K\"apyl\"a et al.\ 2013).
With such a multi-cell \mc, we are able to retain all the attractive features of the
flux transport dynamo model.  The phase relation between the toroidal and the
poloidal fields is correctly reproduced.  The observed correlation between the
polar field during a sunspot minimum and the strength of the next cycle is also
reproduced when the diffusivity is high, although a reduced diffusivity diminishes
this correlation.

One of the important processes in the operation of the flux transport dynamo is
the transport of the poloidal field generated near the surface by the 
Babcock--Leighton mechanism to the bottom of the convection zone where the
differential rotation can act on it.  We have taken the diffusivity on the
higher side in the calculations presented in \S~3--4 
and we find that the poloidal field can diffuse from the surface
to the bottom of the convection zone in a few years.  A complicated multi-cell
\mc\ does not get in the way of this process. However, when the diffusivity
is reduced, this transport has to be done by the \mc.  Interestingly, even
in the case of low diffusivity with a multi-cell \mc, we are still able to
get periodic solutions, although the poloidal field within the convection zone
becomes very complicated and the cycle period is
lengthened. A downward turbulent pumping helps in 
reducing the differences between the high and the low diffusivity models.  
There is no concensus at the present time about latitudinal pumping.  However, 
we reproduce the result of Guerrero
\& de Gouveia Dal Pino (2008) that an equatorward pumping at the bottom of the
convection zone can make a flux transport dynamo work even in the absence of
a flow there. Since such equatorward pumping can have a profound effect on the
dynamo, the nature of such pumping needs to be investigated thoroughly through
magnetoconvection simulations.

To sum up, we do not think that the recent claims of an equatorward return
flow at a shallow depth pose a threat to the flux transport dynamo
model.  Especially, we see no reason to give up the attractive scenario that
the strong toroidal field is produced and stored in the stable regions of the
tachocline, from which parts of this toroidal field break away to rise through
the convection zone and produce sunspots. 
%So we do not agree with the model of the dynamo limited to the upper layers of the convection zone proposed by Pipin \& Kosovichev (2013). 
The crucial assumption needed to make the flux transport
dynamo work is an equatorward flow in low latitudes at the bottom of the convection
zone.  At present, we do not have observational data either supporting or contradicting
it. Since the flux transport dynamo has been so successful in explaining so many
aspects of the solar cycle, we expect this assumption of equatorward flow in low 
latitudes at the bottom of the convection zone to be correct and we hope that
future observations will establish it.  Only if future observations show this
assumption to be incorrect, a drastic revision of our current ideas about the
solar dynamo will be needed at that time.

\bigskip

We thank an anonymous referee for valuable comments.
This work was partially supported by the JC Bose Fellowship awarded to ARC by
the Department of Science and Technology, Government of India. 

\section*{Appendix}
\section*{Stream functions for the three-cell and more complicated meridional circulation}

%\subsection*{Meridional circulation with three radially stacked cells:}
To get three radially stacked cells shown in Figure~\ref{fig:radc3}(a), we take the stream function as  
\begin{equation}
\psi = \psi_u + \psi_m + \psi_l
\end{equation}
The stream function which generates the upper cell is given by
\begin{eqnarray}
\psi_u = {\psi_{0u}}\left[1-{\rm erf}\left(\frac{r-0.87R_\odot}{1.5}\right)\right](r - R_{m,u})^{0.3}\nonumber \\
\times\sin \left[ \frac{\pi (r - R_{m,u})}{(R_\odot -R_{m,u})} \right]\{ 1 - e^{- \beta_1 \theta^{\epsilon}}\}\nonumber \\
\times\{1 - e^{\beta_2 (\theta - \pi/2)} \} e^{-((r -r_0)/\Gamma)^2} ~~~~
\end{eqnarray}\\
where the parameters have the following values:
$\beta_1 = 3.5, \beta_2 = 3.3$, $r_0 = (R_\odot - R_b)/3.5$, $\epsilon = 2.0000001$, $\Gamma =3.4 \times 10^{8}$ m, $R_{m,u} = 0.82 R_\odot$.
%***************************************************
The stream function for middle cell is given by
\begin{eqnarray}
\psi_m={\psi_{0m}}\left[1-{\rm erf}\left(\frac{r-0.85R_{m,u}}{1.5}\right)\right](r - R_{m,l})\nonumber \\
\times\sin \left[ \frac{\pi (r - R_{m,l})}{(R_{m,u} -R_{m,l})} \right]\{ 1 - e^{- \beta_1 \theta^{\epsilon}}\}\nonumber \\
\times\{1 - e^{\beta_2 (\theta - \pi/2)} \} e^{-((r -r_0)/\Gamma)^2} ~~~~
\end{eqnarray}\\
where the parameters have the following values:
$\beta_1 = 1.9, \beta_2 = 1.7$, $r_0 = (R_\odot - R_b)/3.5$, $\Gamma =3.4 \times 10^{8}$ m, $R_{m,l} = 0.75 R_\odot$, $R_{m,u}=0.82 R_\odot$.
Finally, the stream function which generates lower cell is
\begin{eqnarray}
\psi_l = {\psi_{0l}}\left[1-{\rm erf}\left(\frac{r-0.75R_{m,l}}{0.8}\right)\right](r - R_p)\nonumber \\
\times\sin \left[ \frac{\pi (r - R_p)}{(R_{m,l} -R_p)} \right]\{ 1 - e^{- \beta_1 \theta^{\epsilon}}\}\nonumber \\
\times\{1 - e^{\beta_2 (\theta - \pi/2)} \} e^{-((r -r_0)/\Gamma)^2} ~~
\end{eqnarray}\\
where the parameters have the following values:
$\beta_1 = 1.5, \beta_2 = 1.3$, $r_0 = (\Rs - R_b)/3.5$, $\Gamma =3.47 \times 10^{8}$ m, $R_p = 0.65 R_\odot$, $R_{m,l} = 0.76 R_\odot$.  
We choose $\psi_{0u}/C$, $\psi_{0m}/C$ and $\psi_{0l}/C$ in such a way that $v_0$ for upper cell, middle cell and lower cell are around $17.0$ m s$^{-1}$, $5.5$ m s$^{-1}$ and $2.0$ m s$^{-1}$ respectively.
%****************************************

In order to get the complicated \mc\ as shown in Figure~\ref{fig:mcomp_cells}, we choose our stream function as given below.
\begin{equation}
\psi = \psi_l + \psi_{lm} + \psi_m +\psi_u + \psi_{uc}
\end{equation}\\
where,$\psi_l, \psi_{lm}, \psi_m, \psi_u$ and $\psi_{uc}$ generate respectively the lower cell, lower middle cells, 
middle cells, the complicated upper cell and the upper corner cell. 
%Expressions of $\psi_l, \psi_{lm}$ and $\psi_m$ are similar as Eq.$\ref{eq:psi}$ multiplied by $\cos(q)$ factor to get latitudinal division of %cells and an $\theta$ dependent error function to include latitudinal dependence. The upper complicated cell {$\psi_u$} is given below
To give an idea about the kind of stream function we use in order to get a complicated cell, we write down the
stream function $\psi_u$ for the most complicated upper cell:
\begin{eqnarray}
\label{eq:com}
\psi_u = {\psi_{0u}}\left[1+{\rm erf}\left(\frac{r-R_c}{0.02\Rs}\right)\right]\sin \left[ \frac{\pi (r - R_p)}{(\Rs -R_p)} \right]\nonumber \\
\times(r - R_p)\{ 1 - e^{- \beta_1 \theta^{\epsilon}}\}\{1 - e^{\beta_2 (\theta - \pi/2)} \} e^{-((r -r_0)/\Gamma)^2}, ~~~~~
\end{eqnarray}
where 
$$
R_c=\frac{1}{2}\left[1+{\rm erf}\left(\frac{\theta -\pi/24}{\pi/7}\right)\right]{\times}0.95\Rs 
$$
and the parameters  have the following values:

$\beta_1 = 0.45, \beta_2 = 1.3 $, $r_0 = (R_\odot - R_b)/3.5$, $\Gamma =3.1 \times 10^{8}$ m, $R_p = 0.65 R_\odot$. Here do not write down the other stream functions, which are constructed along similar lines.
%***************************************************

 %References*******************************************************

\end{document}